\documentclass[preprint]{aastex}

\slugcomment{Accepted by the Astronomical Journal, 2006 Oct. 11}

\shortauthors{Boboltz et al.}
\shorttitle{Radio Star Astrometry}

\begin{document}

\title{VLA+PT Astrometry of 46 Radio Stars} 
\author{D. A. Boboltz\altaffilmark{1}, 
A. L. Fey\altaffilmark{1}, 
W. K. Puatua\altaffilmark{1},
N. Zacharias\altaffilmark{1}, 
M. J. Claussen\altaffilmark{2}, \\
K. J. Johnston\altaffilmark{1} and 
R. A. Gaume\altaffilmark{1}}

\altaffiltext{1}{U.S. Naval Observatory,
3450 Massachusetts Ave., NW, Washington, DC 20392-5420}
\altaffiltext{2}{National Radio Astronomy Observatory,
P.O. Box O, Socorro, NM 87801}

\begin{abstract}  

We have used the Very Large Array (VLA), linked with the Pie Town Very Long Baseline
Array antenna, to determine astrometric positions of 46 radio stars in the 
International Celestial Reference Frame (ICRF).  Positions were obtained 
in the ICRF directly through phase referencing of the stars to nearby 
ICRF quasars whose positions are accurate at the 0.25 mas level.  Radio 
star positions are estimated to be accurate at the 10~mas level, with position 
errors approaching a few milli-arcseconds for some of the stars observed.
Our measured positions were combined with previous measurements 
taken from as early as 1978 to obtain proper motion estimates for all 46 stars
with average uncertainties of  $\approx$1.7~mas~yr$^{-1}$.  We compared 
our radio star positions and proper motions with the {\em Hipparcos} Catalogue data, 
and find consistency in the reference frames produced by each data set on the 
1$\sigma$ level, with errors of $\sim$2.7 mas per axis for the reference frame 
orientation angles at our mean epoch of 2003.78.  No significant spin is found 
between our radio data frame and the {\em Hipparcos} Celestial Reference Frame (HCRF)
with largest rotation rates of +0.55 and $-$0.41~mas~yr$^{-1}$ around the $x$ and $z$ axes,
respectively, with 1$\sigma$ errors of 0.36 mas~yr$^{-1}$.  Thus, our results are 
consistent with a non-rotating {\em Hipparcos} frame with respect to the ICRF.

\end{abstract}


\section{INTRODUCTION}

The current realization of the International Celestial Reference Frame (ICRF) is 
defined by the positions of 212 extragalactic objects derived from Very Long 
Baseline Interferometry (VLBI) observations \citep{MA:98,IERS:99,FEY:04}.  
This VLBI realization of the ICRF is currently the International Astronomical 
Union (IAU) sanctioned fundamental celestial reference frame.  At optical 
wavelengths, the {\em Hipparcos} Catalogue \citep{PERRYMAN:97} now serves 
as the primary realization of the celestial reference frame.  The link between the 
{\em Hipparcos} Catalogue  and the ICRF was accomplished through a variety of 
ground-based and space-based efforts \citep{KOVAL:97} with the highest weight 
given to VLBI observations of 12 radio stars by \cite{LPJPRTRG:99}.  The standard 
error of the alignment was estimated to be 0.6~mas at epoch 1991.25, with an 
estimated error in the system rotation of 0.25~mas~yr$^{-1}$ per axis \citep{KOVAL:97}.  

At the epoch of our most recent radio star observations (2004.80) the formal error associated 
with the {\em Hipparcos}-ICRF frame link is estimated to be $\sim$3.4~mas.  Due to
errors in the proper motions, the random position errors of individual {\em Hipparcos} stars
increased from $\sim$1~mas in 1991 to $\sim$12~mas at the time of our most recent 
observations.  Such uncertainties in the frame rotation and the astrometry of 
individual sources can combine to seriously limit the ability to align high-resolution 
multi-wavelength data on a particular source thus restricting the astrophysical 
interpretation of potentially interesting objects.

In this article, we present X-band radio observations of 46 radio stars 
using the Very Large Array (VLA) in A configuration linked by fiber optic transmission 
line to the Very Long Baseline Array (VLBA) antenna located in Pie Town, New Mexico. 
Both the VLA and VLBA are maintained and operated by the National Radio Astronomy 
Observatory (NRAO)\footnote{The National Radio Astronomy Observatory is a facility 
of the National Science Foundation operated under cooperative agreement by Associated 
Universities, Inc.}. The VLA plus Pie Town (VLA+PT) link \citep{CBSU:99} is a valuable tool for 
radio star astrometry because it provides the high sensitivity of the VLA with nearly twice 
the resolution of the VLA A-configuration alone for high declination sources.  

The work described herein represents a continuation of a long-term program 
(since 1978) to obtain accurate astrometric radio positions and proper 
motions for $\sim$50 radio stars that can be used to connect the current ICRF to 
future astrometric satellite (e.g. {\it Gaia}; SIM PlanetQuest) reference frames.
These stars were originally selected to be observable at both optical
and radio wavelengths, with detectability in the radio being the primary limitation.
Quiescent radio flux densities are on the order of 1--10~mJy, with occasional flares 
of emission $>$100~mJy.  All of the stars have been observed with {\em Hipparcos} and 
have spectral types ranging from A to M and visual magnitudes ranging from 0.58 to 10.80.
Many of the stars are RS CVn and Algol-type binary systems.  Early observations 
in the program were used to connect the radio frame to the FK4 optical reference frame 
\citep{JDFW:85}, while later observations were used to link the radio-based ICRF to 
the {\em Hipparcos} optical reference frame \citep{JDG:03}. 

The astrometric positions derived from the three epochs of VLA+PT observations are 
combined with previous VLA \citep{JDFW:85, JDG:03}, VLA+PT \citep{BFJCdZG:03},
and Multi-Element Radio Linked Interferometer Network (MERLIN) \citep{FBGJGT:06} 
positions to determine updated proper motions, $\mu_{\alpha \cos\delta}$ and 
$\mu_{\delta}$, for all 46 sources.  Position and proper-motion results obtained for 
the 46 stars are compared with the corresponding {\em Hipparcos} values as a 
measure of the accuracy of our results.  Finally, position and proper motion differences 
relative to the {\em Hipparcos} values are computed in order to determine the current
(epoch 2004) spin alignment of the {\em Hipparcos} frame with respect to the ICRF.

\section{OBSERVATIONS AND REDUCTION \label{OBS}}

\begin{figure}[hbt]
\plotone{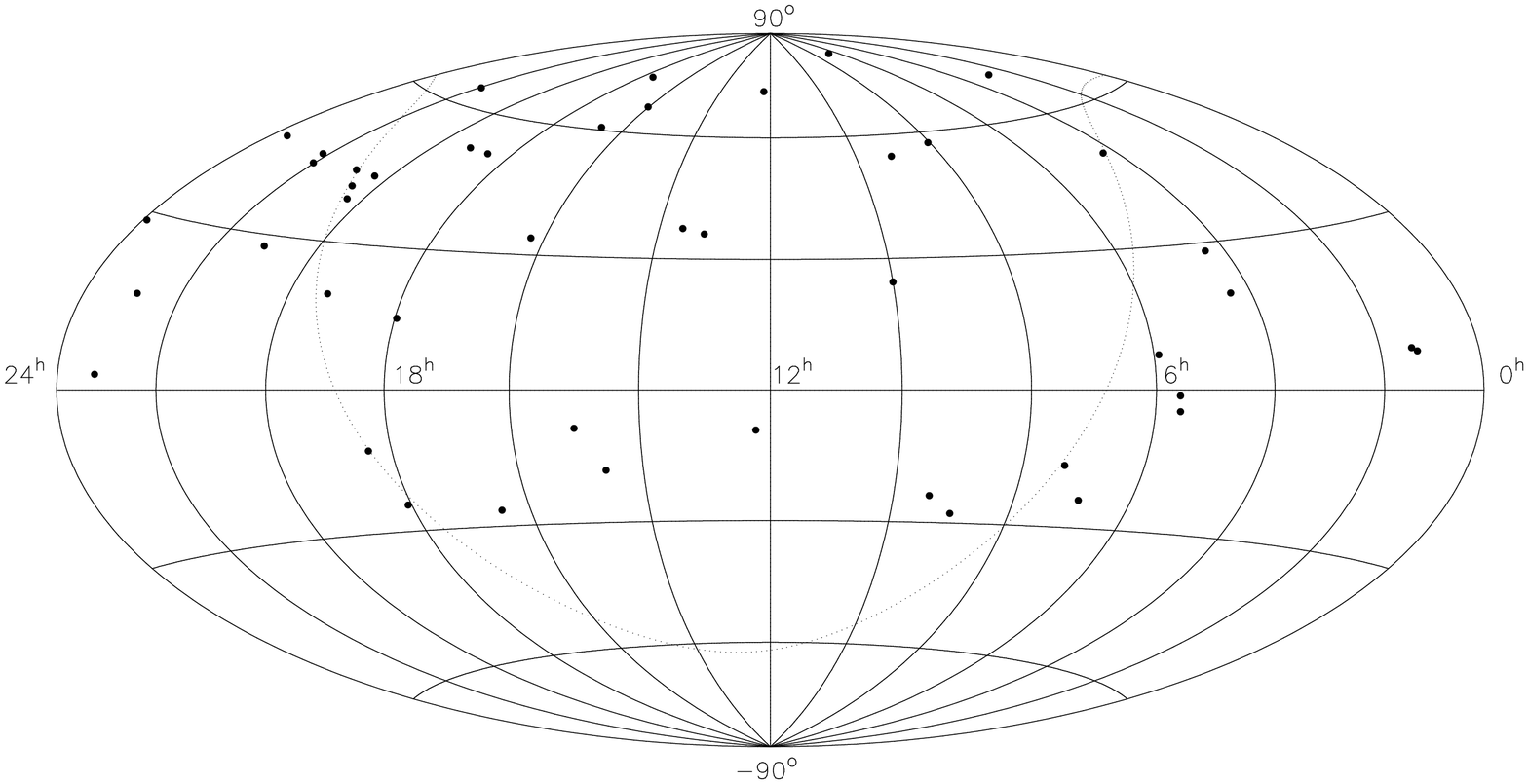}
\figcaption{Distribution of the 46 observed radio stars plotted on an Aitoff equal-area
projection of the celestial sphere.  The dotted line represents the Galactic equator. 
\label{AITOFF}}
\end{figure}

The VLA+PT radio observations occurred over three epochs 2003 June 6$-$7, 
2003 September 9$-$10 and 2004 October 18$-$19.  For the first two epochs, designated
experiments AF399a and AF399b, observations occurred over a 24-hour period 
with 24 and 26 stars observed in each session respectively.  
The third epoch was observed over a 10-hour period in which 10 stars 
not detected in the previous two sessions were re-observed.  Data for all three 
epochs were recorded in dual circular polarization using two adjacent 
50-MHz bands centered on rest frequencies of 8460.1 MHz and 8510.1 MHz 
respectively.  The sky distribution of the 46 radio stars detected is shown in 
Figure \ref{AITOFF}.  

Observations were conducted in a phase-referencing mode by rapidly switching
between the star and a nearby ICRF reference source.  Listed
in Table \ref{SOURCES} are the radio star targets along with their associated 
ICRF calibrators.   Also shown in the table are the ICRF positions for each
calibrator, the ICRF category, and the separation in degrees between the target 
and the reference source.  Positions for the ICRF reference sources are estimated to be 
accurate to the 0.25 mas level.  For the first two epochs, typical target/calibrator scans 
lasted 8 minutes with a 2 minute cycle time (90 seconds on the star and 30 seconds 
on the calibrator) for approximately 4 cycles per scan.  For the third epoch, the cycle times 
were increased to 3 minutes (140 seconds on the star and 40 seconds on the calibrator)
with scans lasting 12 minutes, again resulting in 4 cycles per scan.  Over the 
course of an experiment, 5-8 scans were recorded for each target/calibrator pair
over a wide range of hour angles.  In addition, periodic scans on the 
source 3C48 were performed for the purpose of absolute flux density calibration.

Data were calibrated using the standard routines within the Astronomical 
Image Processing System (AIPS).  The absolute flux density scale was established 
using the values calculated by AIPS for 3C48 with the proper $uv$ restrictions 
applied.  Phase calibration was accomplished through transfer of the phases 
from the reference source to the target source data.  From the calibrated data, 
images were produced for each scan on each target for a total of up to 8 images
per star per epoch of observations.   Average root-mean-square ({\em rms}) noise 
levels in the CLEANed images from the individual scans were 
0.1, 0.09, and 0.04 mJy/beam for AF399a, AF399b, and AJ315, respectively.  
Recall that scan times were increased in 
experiment AJ315 to increase the probability of detecting previously undetected 
stars from AF399a-b.  In addition, a summed image of each star was produced 
which included data from all scans on the source.  Two-dimensional (2-D) 
Gaussian functions were fit to the 
emission in the images using the AIPS task JMFIT.  For the three experiments, 
detection rates were 19 out of 24 stars (79\%) in AF399a, 21 of 26 (81\%) in AF399b, 
and 7 of 11 (64\%) in AJ315.   For comparison, the detection rate for our VLA+PT 
radio-star pilot study \citep{BFJCdZG:03} was 19 out of 19 stars (100\%), however, 
there we purposely tried to select radio stars with high flux densities based upon 
previous observations.

\section{RESULTS AND DISCUSSION}

\subsection{Source Positions \label{POS}}

Final estimation of source positions and uncertainties was performed outside of AIPS 
using the results of the 2-D Gaussian fits to the images produced from the observations.  
Table \ref{POSITIONS} lists the source positions and associated uncertainties 
determined for the 46 stars detected over the three epochs.  Note that the star
HD 193793 appears twice since it was observed in two experiments, 
AF399a and AJ315.  Because the 
radio stars were directly referenced to ICRF quasar calibrators using the 
phase-referencing technique, the positions listed in Table~\ref{POSITIONS}
are given directly in the ICRF.   Also denoted in 
the last two columns of Table~\ref{POSITIONS} are the epoch of observation
and the number of successful/total observations (scans) which were used in the estimation of 
position uncertainties for each source.  Final positions reported in Table~\ref{POSITIONS} 
are simply the JMFIT least squares position estimates from the summed image of each 
star.  The 1$\sigma$ position uncertainties listed in the table were estimated using a procedure 
similar to that described in \cite{FBGJGT:06} and summarized below, depending 
on whether the source was detected in one or more individual scans. 

If a given star was detected in more than one scan, then an {\em rms} scatter 
in the JMFIT scan-based positions weighted by the JMFIT formal errors was computed. 
The uncertainty in the position reported for each star is then the root-sum-square 
({\em rss}) of this {\em wrms} position scatter and the value of the JMFIT least-squares formal
uncertainty from the fit to the summed image of the source.  The addition of the {\em wrms} position 
scatter was meant to conservatively account for possible systematic errors introduced 
into the positions by factors such as the variable troposphere. The position uncertainties 
listed in Table~\ref{POSITIONS} represent the resulting {\em rss} values for sources 
detected in more than one observation. 

If the source was detected in only a single scan, then the reported position uncertainties in 
Table  \ref{POSITIONS} were estimated by taking the {\em rss} of the JMFIT formal position 
error from the summed image, and the average value of the {\em wrms} position scatter 
for all sources with multiple observations in the particular epoch in which the star was observed.  
There were only three such sources, UV~Psc (HIP~5980), SV~Cam (HIP~32015) and 
HR~7275 (HIP~94013), in which the average 
scatter had to be used, one source in each experiment.  For the three experiments,
AF399a, AF399b and AJ315, the average {\em wrms} values of position scatter for stars 
detected in more than one observation were 7.3, 12.1 and 6.3 mas in $\alpha \cos\delta$ 
and 9.4, 11.1, and 9.9 mas in $\delta$, respectively.  Again, this {\em rss} step was meant to 
conservatively account for possible systematic errors in the measured positions. 

Table~\ref{POS_ERRS_TAB} compares the uncertainties in our VLA+PT positions 
with the corresponding {\em Hipparcos} uncertainties and lists the {\em rss} 
combined uncertainties for each star.   The {\em Hipparcos} uncertainties have been
updated to the epoch of our observations using the reported {\em Hipparcos} proper 
motion errors.   Our errors compare favorably with the {\em Hipparcos} position 
errors updated to our epoch.  The average (median) position uncertainties for all 
46 stars detected in the VLA+PT observations is 10.2~mas (9.2~mas) in $\alpha \cos\delta$ 
and 11.5~mas (11.3~mas) in $\delta$.  The average (median) {\em Hipparcos} position 
uncertainties are slightly larger in $\alpha \cos\delta$ at 12.0~mas (10.3~mas) and slightly
smaller in $\delta$ at 10.3~mas (8.8~mas).

\begin{figure}[hbt]
\plotone{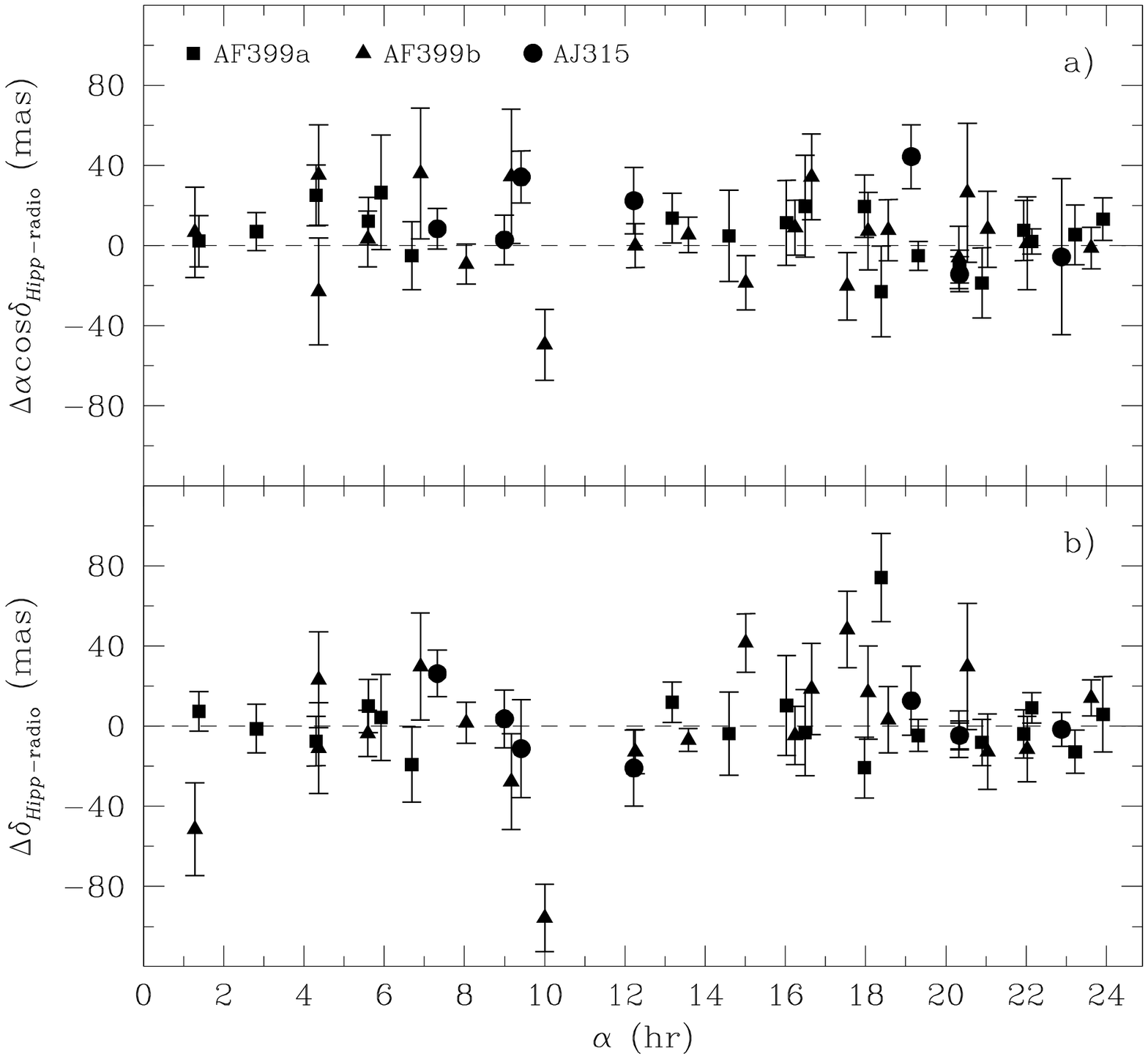}
\figcaption{Differences between the {\em Hipparcos} positions updated to the epoch of our 
observations, and our VLA+PT measured positions as a function of source right 
ascension $\alpha$ for the 46 radio stars observed.  Differences in $\alpha \cos\delta$ 
are plotted in (a) and differences in $\delta$ are plotted in (b). Error bars are 
are the {\em rss} combined uncertainties listed in Table~\ref{POS_ERRS_TAB}. \label{HIPVLARA}}
\end{figure}

\begin{figure}[hbt]
\plotone{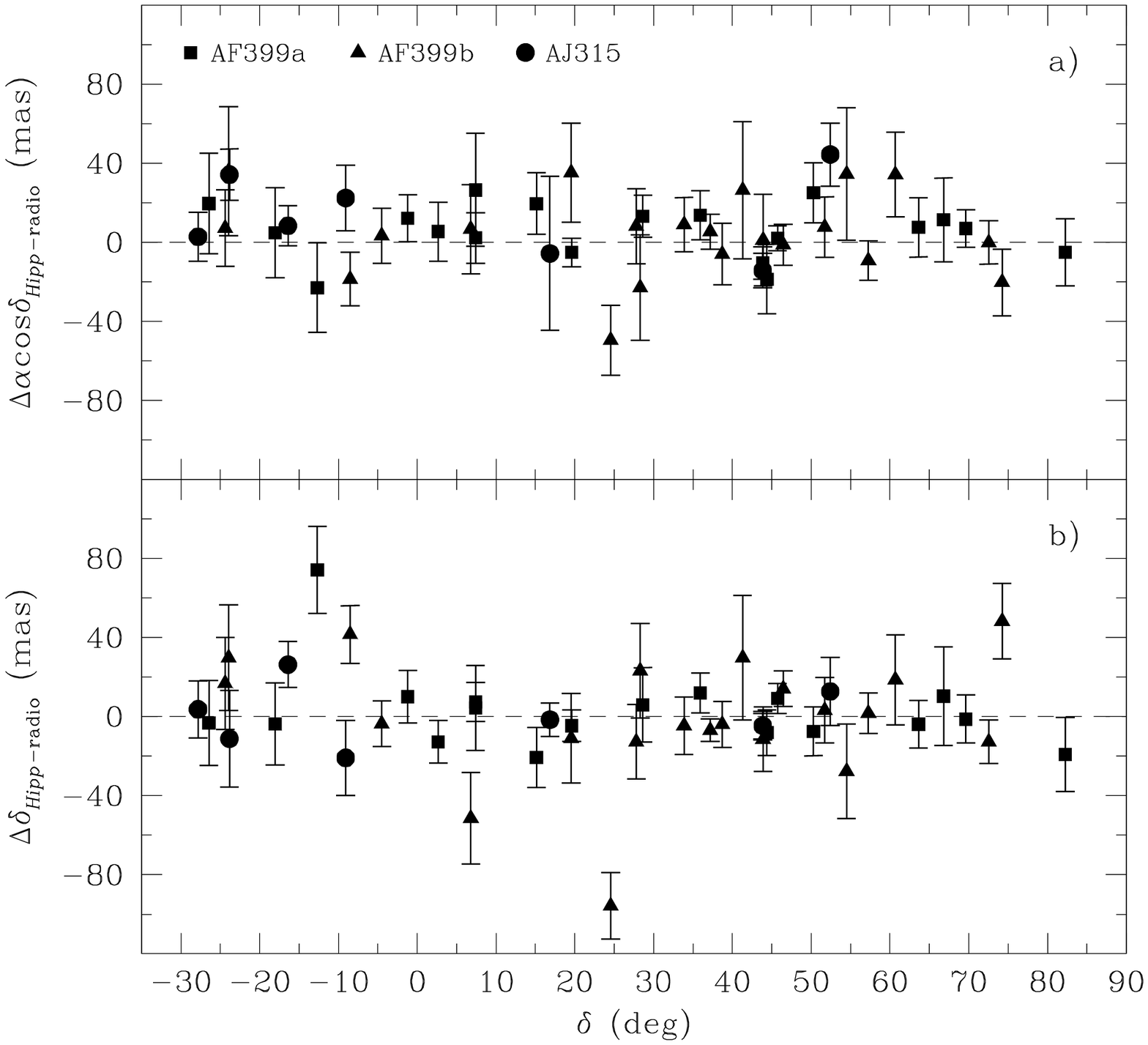}
\figcaption{Differences between the {\em Hipparcos} positions updated to the epoch of our 
observations, and our VLA+PT measured positions as a function of source declination
$\delta$ for the 46 radio stars observed.  Differences in $\alpha \cos\delta$ 
are plotted in (a) and differences in $\delta$ are plotted in (b). Error bars are 
are the {\em rss} combined uncertainties listed in Table~\ref{POS_ERRS_TAB}. \label{HIPVLADEC}}
\end{figure}

Listed in the last two columns of Table~\ref{POS_ERRS_TAB} 
are the offsets between our our VLA+PT positions and the {\em Hipparcos} positions 
updated to the epoch of our observations.   These offsets, $\Delta_{Hipp.-{\rm radio}}$,
are also shown as a function of right ascension in Figure~\ref{HIPVLARA}
and as a function of declination in Figure~\ref{HIPVLADEC}.  Stars observed in the three 
different experiments are represented by different symbols.  Error bars are the 
combined uncertainties reported in columns 7 and 8 of Table~\ref{POS_ERRS_TAB}.
The un-weighted mean offsets between our positions and the 
updated {\em Hipparcos} positions are 6.1~mas in $\Delta\alpha \cos\delta$ and 
0.7~mas in $\Delta\delta$ with standard deviations ($\sigma$) of 18.6~mas and 
25.1~mas in $\Delta\alpha \cos\delta$ and $\Delta\delta$, respectively.
Un-weighted standard errors of the means ($\sigma_M$) are then 2.7~mas 
in $\Delta\alpha \cos\delta$ and 3.7~mas in $\Delta\delta$. 
Similarly, the mean offsets between the VLA+PT and {\em Hipparcos} 
positions, weighted by the square of the {\em rss} combined uncertainties,
are 2.3~mas and $-$0.7~mas with weighted {\em rms} 
errors of 14.3~mas and 17.4~mas in $\Delta\alpha \cos\delta$ and 
$\Delta\delta$, respectively. 

The un-weighted average (median) arc length between our measurements and 
the {\em Hipparcos} positions is 24.2~mas (15.7~mas) with a standard 
deviation of 20.5~mas.  There are 
two stars for which the arc length is greater than 75~mas; DH~Leo (HIP~49018)  
and FR~Sct (HIP~90115).  For both stars, the declination offset is the dominant source of 
the difference with {\em Hipparcos}, however, neither source has an unusually large 
uncertainty in declination, 13.6~mas for DH~Leo and 17.0~mas for FR~Sct.
In addition, neither source has a particularly large proper motion in declination, 
$-$31.8~mas~yr$^{-1}$ and $-$2.9~mas~yr$^{-1}$ for DH~Leo and FR~Sct,
respectively.   

DH~Leo is an RS CVn binary as are many of the radio 
stars on our list.  The system is flagged as a component solution in the {\em Hipparcos} 
Double/Multiple Systems Annex \citep{PERRYMAN:97}.  The annex lists a tertiary component 
having a separation of 220$\pm$20~mas with respect to DH~Leo at a position angle 
of 46~degrees at epoch 1991.25.  DH~Leo is also listed in the {\it Fourth Catalog 
of Interferometric Measurements of Binary Stars} \citep{HMWM:01} as multiple 
system CHARA 145.  The catalog lists eight measurements of the component 
separations made with speckle interferometry from epoch 1989.2271 through 1994.2209.  
Over this 5-yr period, the components moved through angles from 38 to 28 degrees and relative 
separations of 216~mas to 283~mas.   It is therefore possible that the 96~mas offset 
in declination between our radio position and the {\em Hipparcos} position updated to 
our epoch is consistent with the orbital motion within the system.

FR~Sct, on the other hand, is a single pulsating variable star.  The {\em Hipparcos}
Catalogue solution contains no entry in the Double/Multiple Systems Annex \citep{PERRYMAN:97}.   
In addition, there are no entries for FR Sct in the {\em Washington Double Star Catalog} 
\citep{MWHDW:01} or the  {\em Fourth Catalog of Interferometric Measurements of Binary Stars} 
\citep{HMWM:01}.  Therefore, the 74~mas offset in declination between our
radio position and the corresponding {\em Hipparcos} position cannot, as yet, 
be explained as motion due to a secondary component.  It may be that FR~Sct
is a good candidate for future speckle interferometry observations in light of the 
offset we have found. 

\subsection{Source Proper Motions \label{PROP}}

The positions of the 46 radio stars from our VLA+PT observations were 
combined with previous VLA \citep{JDFW:85, JDG:03}, VLA+PT \citep{BFJCdZG:03},
and MERLIN \citep{FBGJGT:06} positions to determine updated proper motions, 
$\mu_{\alpha \cos\delta}$ and $\mu_{\delta}$, for all 46 sources.  
Although the data cover a long time range, 1978--2004, the sampling is not 
sufficient to enable the determination of source parallaxes.  We therefore 
used the {\em Hipparcos} values to remove the effects of parallax in our computed 
proper motions for all 46 stars.  Source proper motions were computed using a 
linear least-squares fit to the data weighted by the position errors for each observation.  
Position errors for the previous VLA-only observations were estimated to be 30~mas in both 
$\alpha \cos\delta$ and $\delta$ \citep{JDG:03} and we have adopted these values.
Position errors for previous VLA+PT and MERLIN observations are reported in 
\cite{BFJCdZG:03} and \cite{FBGJGT:06}, respectively.   The proper motions derived from 
the combined data are listed in Table \ref{PROP_MOT_TAB}.  
Also reported are the number of positions used to determine the proper motion 
($N_{\rm pos}$)
and the total time spanned between the earliest position measurement and the most 
recent measurement ($\Delta \tau$).   There are three stars, T~Tau~N (HIP~20390),
HD~199178 (HIP~100287) and IM~Peg (HIP~112997), for which the time 
baseline is short, $\Delta \tau < 4$~yr.  These three stars 
are recent additions to our observing list and were not
part of the original VLA radio-star program \citep{JDFW:85, JDG:03}.

\begin{figure}[hbt]
\plotone{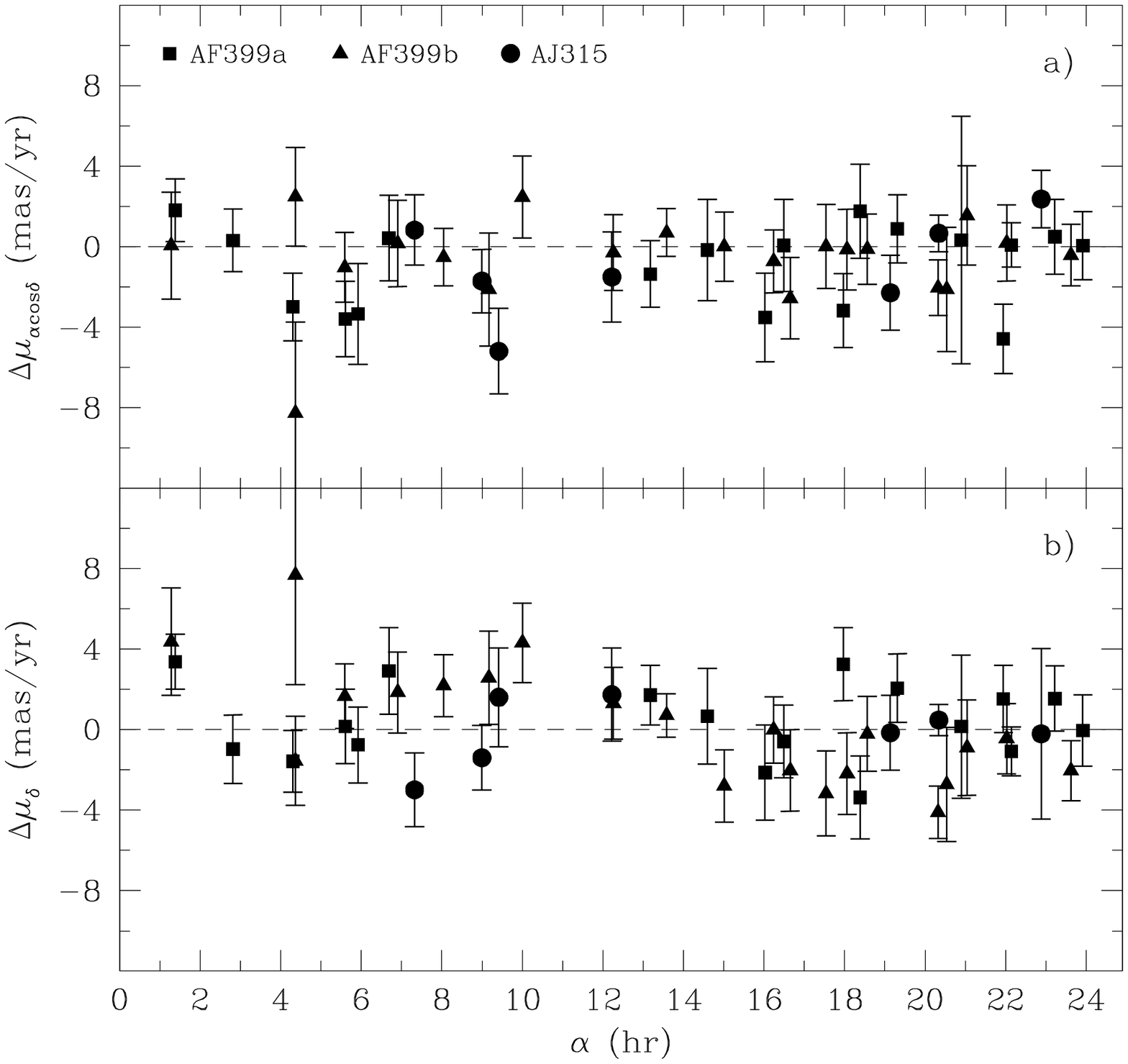}
\figcaption{Differences between the {\em Hipparcos} proper motions and 
our radio derived proper motions as a function of source right 
ascension $\alpha$ for the 46 radio stars observed.  Differences in $\mu_{\alpha \cos\delta}$ 
are plotted in (a) and differences in $\mu_{\delta}$ are plotted in (b). Error bars are 
are the {\em rss} combined uncertainties listed in Table~\ref{PM_ERRS_TAB}. \label{HIPVLA_PM_RA}}
\end{figure}

\begin{figure}[hbt]
\plotone{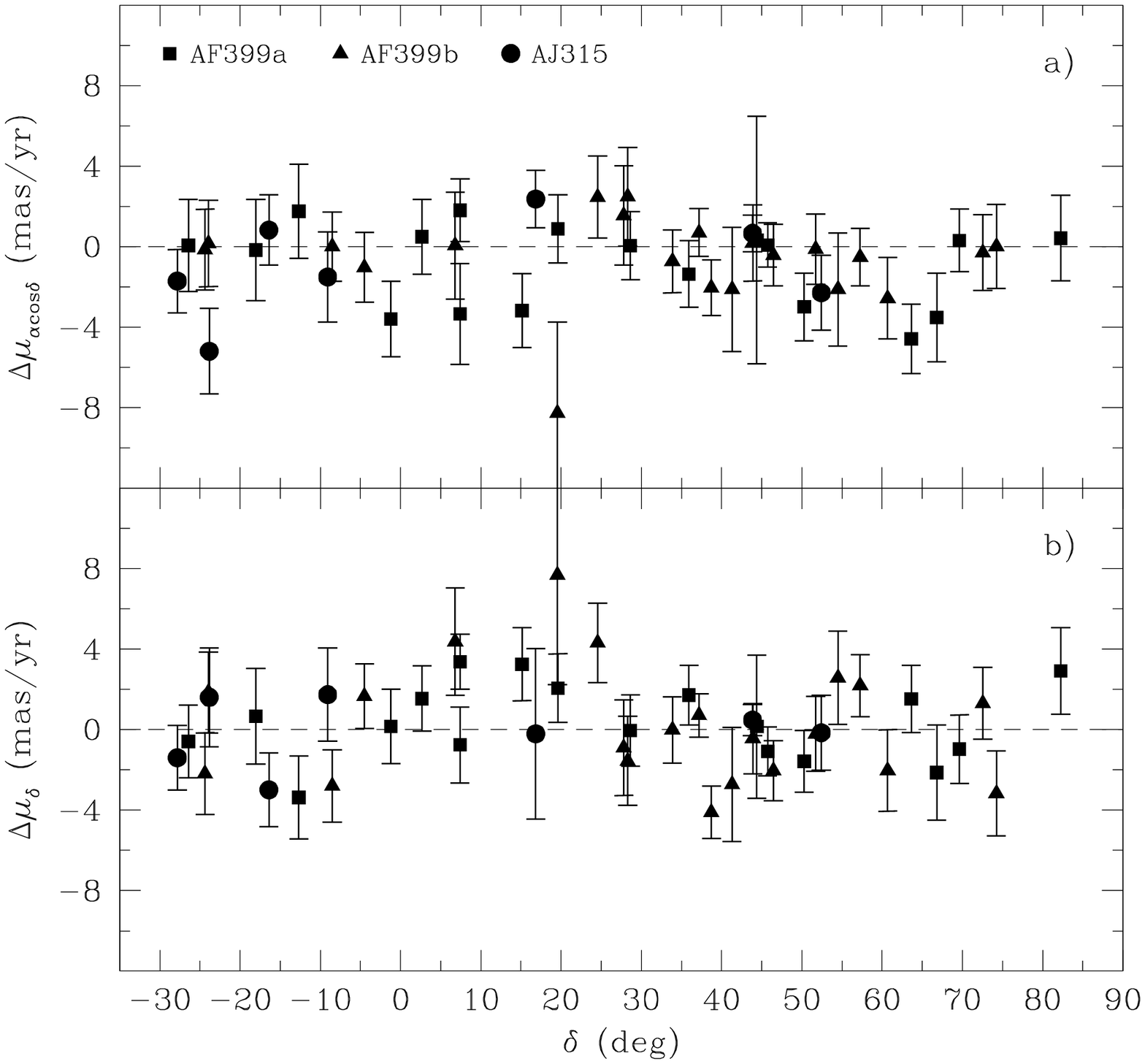}
\figcaption{Differences between the {\em Hipparcos} proper motions and 
our radio derived proper motions as a function of source right 
ascension $\alpha$ for the 46 radio stars observed.  Differences in $\mu_{\alpha \cos\delta}$ 
are plotted in (a) and differences in $\mu_{\delta}$ are plotted in (b). Error bars are 
are the {\em rss} combined uncertainties listed in Table~\ref{PM_ERRS_TAB}. \label{HIPVLA_PM_DEC}}
\end{figure}

\begin{figure}[hbt]
\plotone{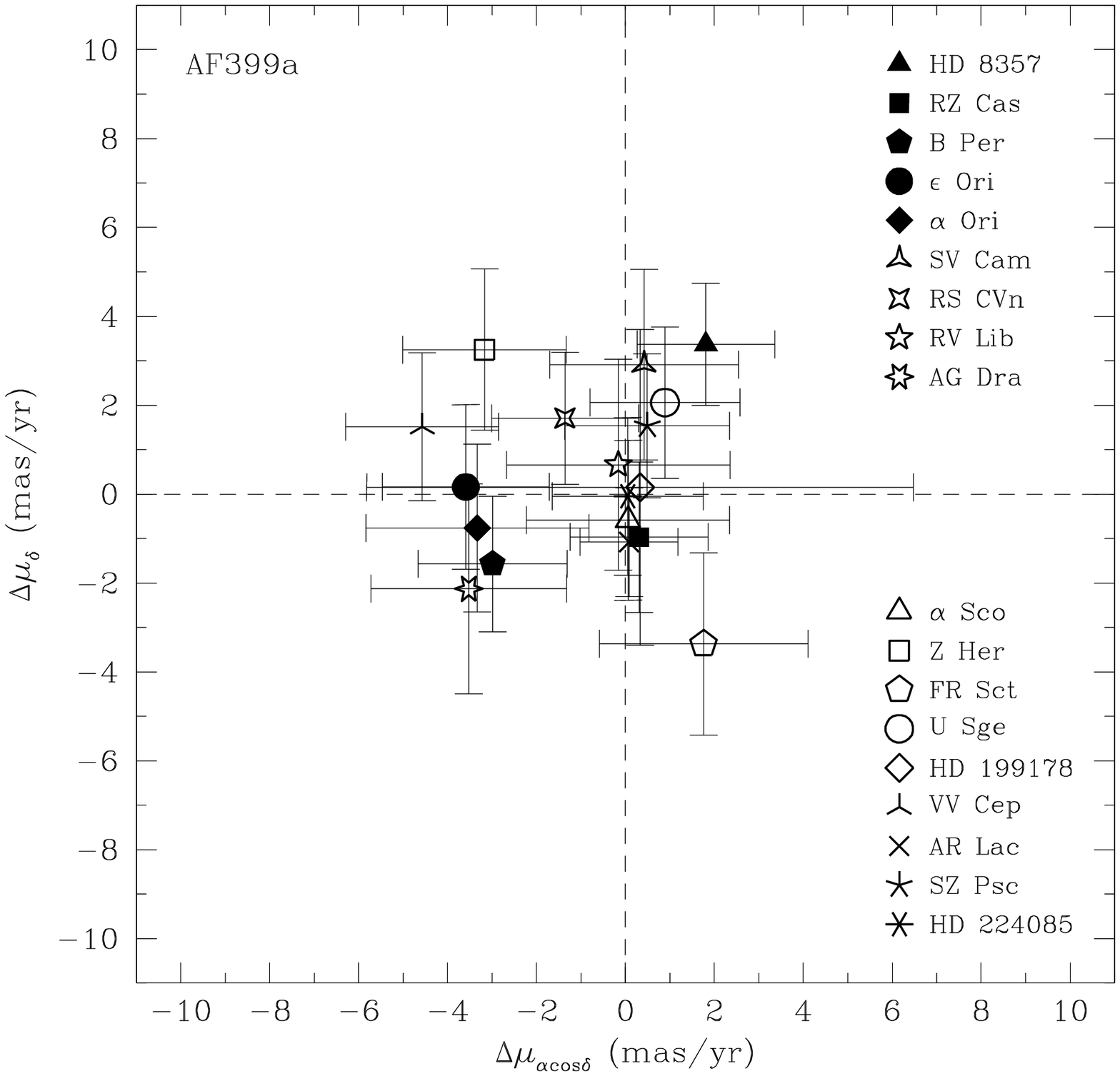}
\figcaption{Offsets between the {\em Hipparcos} and radio proper motions, 
$\Delta\mu_{\alpha \cos\delta}$ vs. $\Delta\mu_{\delta}$, for the stars observed in 
VLA+PT experiment AF399a.  Error bars are are the {\em rss} combined uncertainties 
listed in Table~\ref{PM_ERRS_TAB}. \label{PROP_MOT_AF399a}}
 \end{figure}

\begin{figure}[hbt]
\plotone{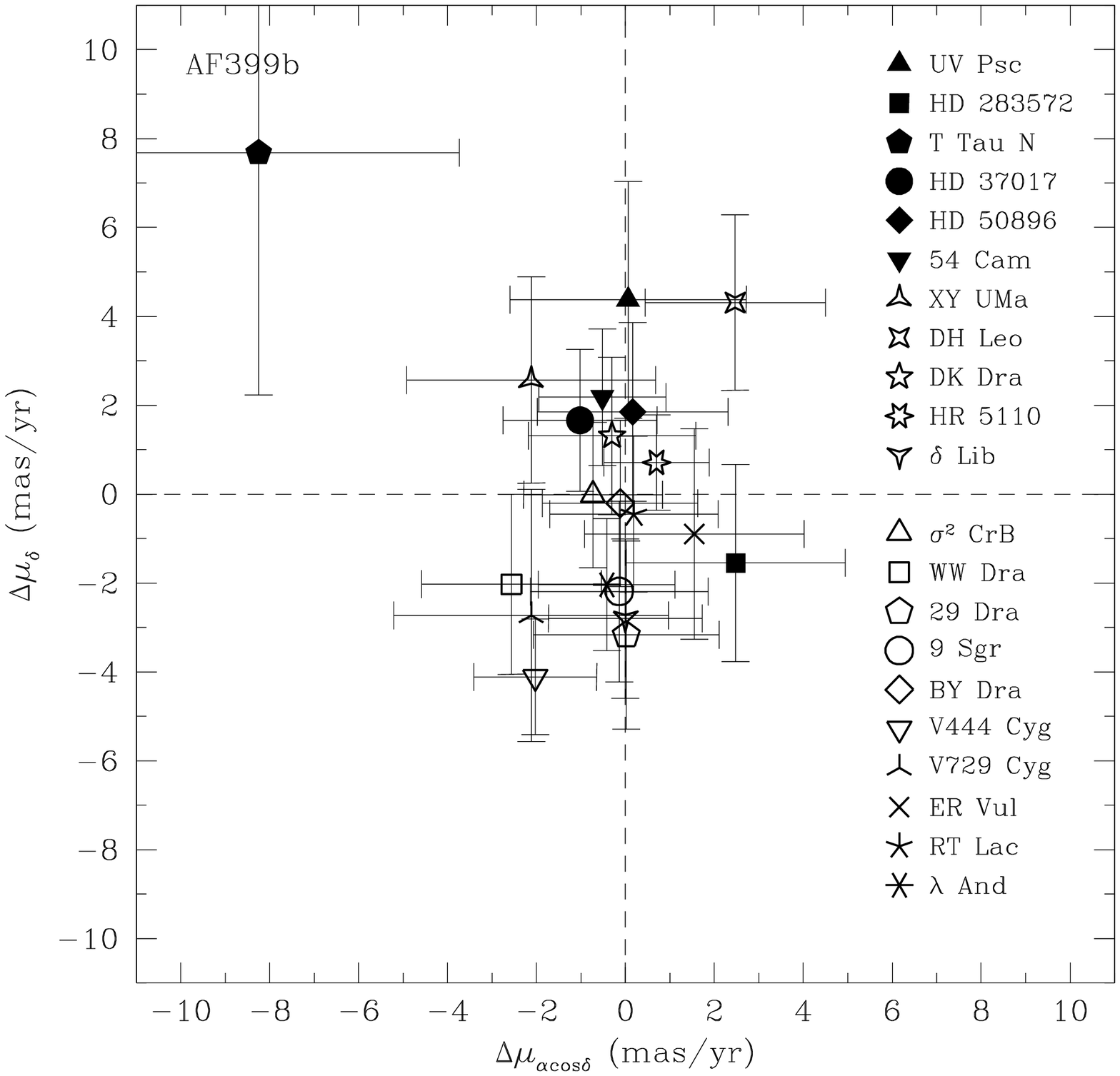}
\figcaption{Offsets between the {\em Hipparcos} and radio proper motions, 
$\Delta\mu_{\alpha \cos\delta}$ vs. $\Delta\mu_{\delta}$, for the stars observed in 
VLA+PT experiment AF399b.  Error bars are are the {\em rss} combined uncertainties 
listed in Table~\ref{PM_ERRS_TAB}. \label{PROP_MOT_AF399b}}
 \end{figure}

\begin{figure}[hbt]
\plotone{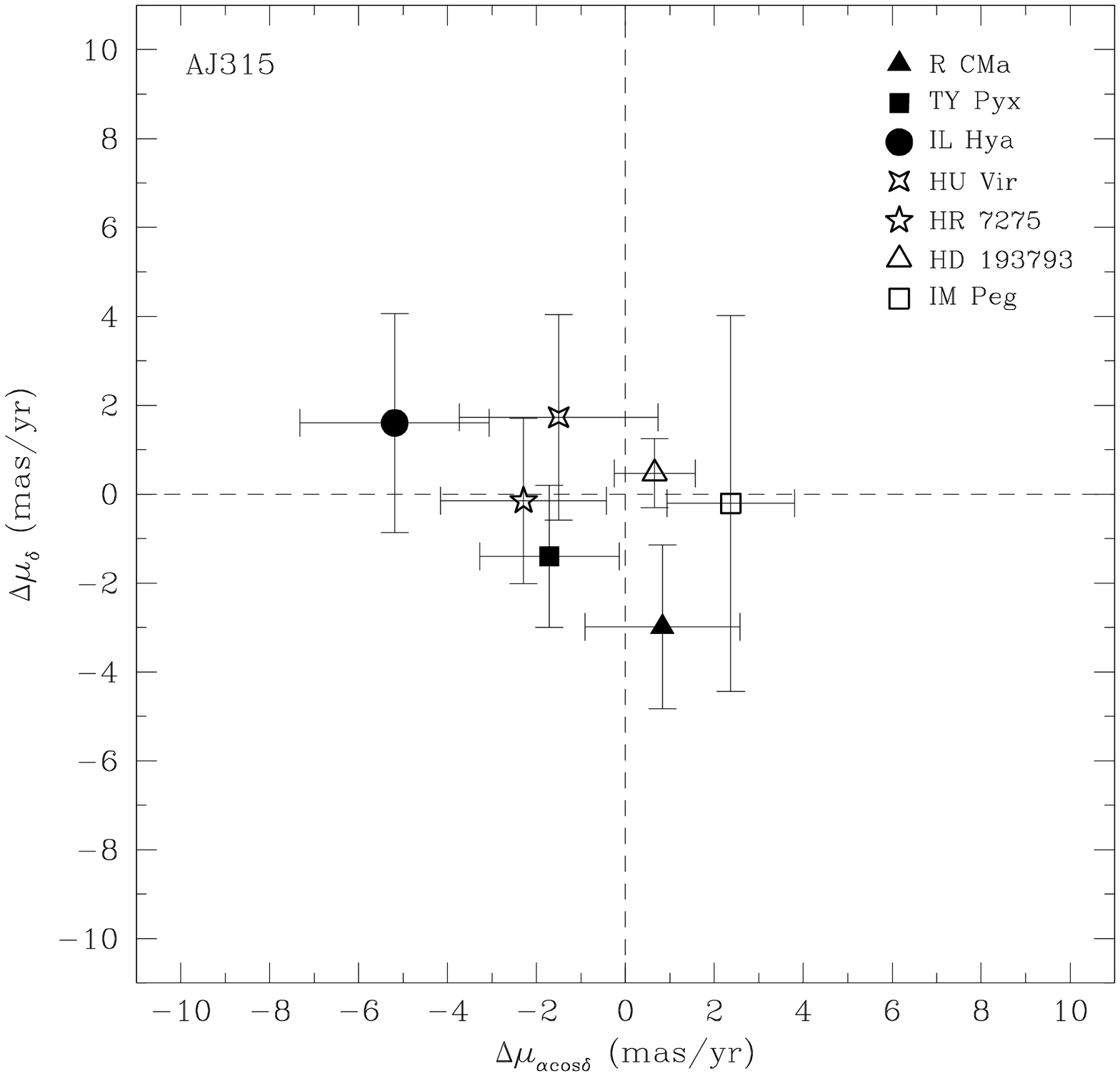}
\figcaption{Offsets between the {\em Hipparcos} and radio proper motions, 
$\Delta\mu_{\alpha \cos\delta}$ vs. $\Delta\mu_{\delta}$, for the stars observed in 
VLA+PT experiment AJ315.  Error bars are are the {\em rss} combined uncertainties 
listed in Table~\ref{PM_ERRS_TAB}. \label{PROP_MOT_AJ315}}
 \end{figure}

Table~\ref{PM_ERRS_TAB} compares the uncertainties in our radio 
derived proper motions with the corresponding {\em Hipparcos} proper 
motion uncertainties and lists the {\em rss} combined uncertainties for each star. 
Listed in the last two columns of Table~\ref{PM_ERRS_TAB} 
are the differences between our our VLA+PT proper motions and the 
corresponding {\em Hipparcos} values.   These differences are also shown in 
Figures~\ref{HIPVLA_PM_RA}--\ref{PROP_MOT_AJ315}.   
Figures~\ref{HIPVLA_PM_RA} and \ref{HIPVLA_PM_DEC} show the proper motion 
differences $\Delta\mu_{\alpha\cos\delta}$ and $\Delta\mu_{\delta}$, as a function of right 
ascension (Fig.~\ref{HIPVLA_PM_RA}) and declination (Fig.~ \ref{HIPVLA_PM_DEC}), 
respectively.  The three different symbols represent
stars observed in the three corresponding VLA+PT experiments.   Error bars are the 
combined uncertainties reported in columns 7 and 8 of Table~\ref{PM_ERRS_TAB}. 
Average (median) radio derived uncertainties for the 46 stars are 1.74~mas~yr$^{-1}$ 
(1.62~mas~yr$^{-1}$) in $\mu_{\alpha \cos\delta}$ and 1.79~mas~yr$^{-1}$ 
(1.65~mas~yr$^{-1}$) in $\mu_{\delta}$.  If we exclude the three stars for
which $\Delta \tau < 4$~yr, these values drop slightly to 1.59~mas~yr$^{-1}$ 
(1.62~mas~yr$^{-1}$) in $\mu_{\alpha \cos\delta}$ and 1.62~mas~yr$^{-1}$ 
(1.64~mas~yr$^{-1}$) in $\mu_{\delta}$.
For comparison, the average {\em Hipparcos} proper motion uncertainties are 
0.98~mas~yr$^{-1}$ (0.85~mas~yr$^{-1}$) in $\mu_{\alpha \cos\delta}$ and 
0.84~mas~yr$^{-1}$ (0.72~mas~yr$^{-1}$) in $\mu_{\delta}$ for the same 46 stars.

The un-weighted average differences between our proper motions and the 
{\em Hipparcos} values are $-$0.75~mas~yr$^{-1}$ in 
$\Delta\mu_{\alpha \cos\delta}$ and 0.21~mas~yr$^{-1}$ in $\Delta\mu_{\delta}$ 
with standard deviations ($\sigma$) of 2.17~mas~yr$^{-1}$ and 
2.38~mas~yr$^{-1}$ in $\Delta\mu_{\alpha \cos\delta}$ and $\Delta\mu_{\delta}$, 
respectively.  Un-weighted standard errors of the means ($\sigma_M$) are thus 
0.35~mas~yr$^{-1}$ in $\Delta\mu_{\alpha \cos\delta}$ and 0.32~mas~yr$^{-1}$ 
in $\Delta\mu_{\delta}$.  Similarly, the mean offsets between the radio 
and {\em Hipparcos} proper motions, weighted by the square of the {\em rss} combined 
uncertainties, are $-$0.57~mas~yr$^{-1}$ and $-$0.15~mas~yr$^{-1}$ with 
weighted {\em rms} errors of 
1.77~mas~yr$^{-1}$ and 2.26~mas~yr$^{-1}$ in $\Delta\mu_{\alpha \cos\delta}$ 
and $\Delta\mu_{\delta}$, respectively. 

Figures~\ref{PROP_MOT_AF399a}--\ref{PROP_MOT_AJ315} plot the differences 
$\Delta\mu_{\delta}$ versus $\Delta\mu_{\alpha \cos\delta}$ for the 
three experiments AF399a, AF399b, and AJ315, respectively.  Again the error 
bars represent the {\em rss} combined uncertainties.  The figures show that many of the 
differences between the radio proper motions and those of {\em Hipparcos} are within the 
1$\sigma$ error bars, with the most obvious exception being the star T~Tau~N in 
Figure~\ref{PROP_MOT_AF399b}.  As mentioned previously, T~Tau~N is one 
of the stars for which the time baseline is short at only 2.75~yr.  In addition, 
T~Tau~N is known to be gravitationally bound to the T~Tau~S binary system 
with a detected acceleration in its motion \citep{JFGCH:04}.  With only
two positions for T~Tau~N covering such a short time period it is impossible to 
fit for any accelerations in the motion of the source with our data alone. 

The two stars mentioned previously as having large declination differences relative to 
{\em Hipparcos}, FR~Sct and DH~Leo, appear in Figures~\ref{PROP_MOT_AF399a}
and \ref{PROP_MOT_AF399b}, respectively.  Although the proper motions in
declination (see Table~\ref{PROP_MOT_TAB}) are not very large, the proper 
motion differences in declination relative to {\em Hipparcos} are fairly large at 
$-$3.35~mas~yr$^{-1}$ for FR~Sct and 4.33~mas~yr$^{-1}$ for DH~Leo.  
Table~\ref{PROP_MOT_TAB} shows that both stars have only two position 
measurements separated by long time intervals between epochs.  It is possible
that the two stars have an acceleration component in declination which is as yet 
undetected in the linear fits to our limited data.  

\subsection{Radio/Optical Frame Alignment \label{FRAMES}}

Our radio star observations are on the ICRF while the
data taken from the {\em Hipparcos} Catalogue  are on the
{\em Hipparcos} Celestial Reference Frame (HCRF).
The {\em Hipparcos} positions used here have been updated to 
the epoch of the individual radio star's mean position using the 
{\em Hipparcos} proper motions.  Following the formulation of \cite{WS:00}
the (optical $-$ radio) position differences are used to
determine the relative reference frame orientation angles
$\epsilon_{x}, \epsilon_{y}, \epsilon_{z}$,
around the $x,y,z$ axes, respectively.

\begin{equation}
( \alpha_{HCRF} - \alpha_{ICRF} ) \cos \delta \ = \
     \epsilon_{x} \sin \delta \cos \alpha 
   + \epsilon_{y} \sin \delta \sin \alpha
   - \epsilon_{z} \cos \delta          
\end{equation}

\begin{equation}
 \delta_{HCRF} - \delta_{ICRF} \ = \
     - \epsilon_{x} \sin \alpha 
     + \epsilon_{y} \cos \alpha 
\end{equation}

Similar formulas are used to obtain the relative spin difference
($\omega_{x}, \omega_{y}, \omega_{z}$)
of the reference frames using the proper motion differences
between the {\em Hipparcos} Catalogue and our data.
The combined {\em rss} formal errors of the {\em Hipparcos} and our
data are used for weighted least-squares adjustments.
The weighted mean epoch of our data is 2003.78 and
results, with the sign conventions from equations 1 and 2, are presented 
in Table~\ref{ROTATION}.  The first two lines of the 
table list the orientation (mas), spin (mas~yr$^{-1}$), and corresponding 
formal errors for each axis using all 46 stars observed.  The HCRF excludes stars
flagged for possible multiplicity in the {\em Hipparcos} Catalogue.
Thirteen out of the 46 radio stars we observed have a 
multiplicity flag in the {\em Hipparcos} Catalogue, thus we have excluded 
them in the second solution presented in Table~\ref{ROTATION}
(labeled as 33 stars).  Finally, two stars out of the remaining 33 non-multiple 
stars showed large post-fit residuals in the 33 star rotation solution.  These 
stars are T Tau N, a known multiple, and RZ Cas.  A third solution 
was produced excluding these two stars and the results are presented
in Table~\ref{ROTATION} (labeled as 31 stars).

Because ground-based catalogs often contain systematic errors especially as 
a function of declination, preliminary solutions for the orientation angles included 
an offset in the declination parameter in addition to the 3 rotation terms.  However, 
these solutions showed the offset term to be insignificant, and the results presented 
in Table~\ref{ROTATION} are based on a model including only the 3 rotation terms.  
The reduced $\chi^2$ was found to be 1.14 for the position orientation solution and 1.10 
for the proper motion spin solution.  This is an indication of small systematic
errors and the addition of an arbitrary {\em rss} error of about 5 mas per 
coordinate per star will bring the $\chi^2$ for the solutions close to 1.0.
This additional error was not included in the solutions presented in 
Table~\ref{ROTATION}.

Updating the {\em Hipparcos}/ICRF frame alignment discussion presented in 
\cite{BFJCdZG:03}, the formal, predicted error on the frame alignment at 
our 2003.78 mean epoch, which is 12.53 years after the mean {\em Hipparcos} 
epoch of 1991.25, is 3.1~mas.  The largest frame orientation angle we find 
is for the $z$-axis with, {\em Hipparcos} $-$ radio = $-$3.2 mas, with a formal 
error of 2.9 mas, thus only a $1\sigma$ (non-)significance.  The orientation 
of the {\em Hipparcos} and ICRF frames are even better for the 
$x$ and $y$ axes at the 2003.78 mean epoch.  For the two alternate solutions 
with 33 and 31 stars, respectively, the frame orientations are slightly larger 
for the $z$ axis than the 46 star solution, and slightly smaller for the $x$ and 
$y$ axes.  All rotation angles are still within the 1$\sigma$ formal errors.  In addition, 
the weighted mean offsets between the {\em Hipparcos} and VLA+PT postions 
mentioned in \S\ref{POS} (2.3 and $-$0.7~mas)  are consistent with the frame orientation 
angles and their formal errors.

With formal errors of the {\em Hipparcos} data increasing over time
and the radio data errors decreasing, it is now appropriate
to look, for the first time, at the derivative of the frame orientation, i.e.~the 
proper motion spin alignment of the frames.
We find formal errors in the spin alignment of only about 0.36~mas~yr$^{-1}$ 
per axis for the 46 star solution.  This is a factor of 2 improvement over our 
previous results \citep{BFJCdZG:03} and is approaching the original {\em Hipparcos}/ICRF 
link error of 0.25 mas~yr$^{-1}$.  Our independent observations show the 
{\em Hipparcos} frame to be non-rotating with respect to the extragalactic ICRF 
with largest rotation rates being +0.55 and $-$0.41~mas~yr$^{-1}$ around 
the $x$ and $z$ axes, respectively.  These rates are consistent with zero on the 
1.6$\sigma$ and 1.1$\sigma$ levels, respectively.   For the 33 and 31 star 
solutions, the formal errors are larger at approximately 0.44~mas~yr$^{-1}$.  The 
rotation rate about the $x$ axis was nominally larger with respect to the 46 star
solution at $\omega_x \approx 0.62$~mas~yr$^{-1}$, while the rotation rate about 
the $z$ axis was slightly smaller at $\omega_z \approx -$0.31~mas~yr$^{-1}$.
All rotation rates are consistent with zero on a 1.5$\sigma$ level or better.
The weighted mean proper motion differences mentioned in \S\ref{PROP} 
($-$0.57 and $-$0.15~mas~yr$^{-1}$) are also consistent with 
these frame rotation rates and their formal errors.

\section{CONCLUSIONS}

We have determined the astrometric positions for 46 radio stars using the 
VLA+PT configuration. The positions presented here, with uncertainties 
on the order of 10 mas or better, are consistent with our earlier VLA+PT results 
\citep{BFJCdZG:03} and represent a factor of three improvement over 
prior VLA-only results \citep{JDFW:85, JDG:03}.  Stellar positions from {\em Hipparcos} 
are degrading with time due to errors in the {\em Hipparcos} proper motions on the order 
of 1~mas~yr$^{-1}$ and due to unmodeled rotations in the frame with respect to the 
extragalactic objects estimated to be 0.25~mas~yr$^{-1}$ per axis.
Taking into account these uncertainties, for many of the stars in our list, 
our VLA+PT positions are better than the corresponding {\em Hipparcos} positions at epoch.  
The proper motions derived from our VLA+PT positions combined with 
previous VLA \citep{JDFW:85, JDG:03}, VLA+PT \citep{BFJCdZG:03},
and MERLIN \citep{FBGJGT:06} positions have errors which are on the 
order of, and in some cases are better than, those obtained from {\em Hipparcos}. 

We have also compared our radio star data with the {\em Hipparcos} Catalogue 
data for positions and proper motions, and find consistency in the reference 
frames produced by each data set on the 1$\sigma$ level.  Errors  
of $\sim$2.8~mas per axis were computed for the reference frame orientation angles
at our mean epoch of 2003.78 and $\sim$0.36 mas~yr$^{-1}$ per axis for 
relative spin between the frames.  Our independent observations show the
{\em Hipparcos} frame to be non-rotating with respect to the extragalactic 
ICRF with largest rotation rates being +0.55 and $-$0.41 mas~yr$^{-1}$ 
around the $x$ and $z$ axes, respectively.
Future papers will reveal if this trend has any significance.
An independent study based on optical images of extragalactic
reference frame sources in combination with dedicated astrograph
observations is in preparation (Zacharias \& Zacharias~2007).
For now the HCRF orientation and spin is consistent with the
ICRF on the 1$\sigma$ level of our observations of 46 radio stars.

\clearpage

\clearpage

\begin{deluxetable}{lrccrrc}
\tabletypesize{\small}
\tablewidth{0pt}
\tablecaption{Observed Radio stars and Corresponding ICRF Calibrator Sources.\label{SOURCES}}
\tablehead{ \colhead{Star} & \colhead{\em Hipparcos} & \colhead{ICRF} &  \colhead{ICRF}
& \colhead{$\alpha$ (J2000)\tablenotemark{b}} & \colhead{$\delta$ (J2000)\tablenotemark{b}} 
& \colhead{Separation} \\
\colhead{Name}    & \colhead{Number} & \colhead{Calibrator} 
& \colhead{Category\tablenotemark{a}} & \colhead{(h m s)} 
& \colhead{($^{\circ}$ $'$ $''$)} & \colhead{($^{\circ}$)} }
\startdata
UV Psc	\dotfill	&	5980	&	0119+041	&	C	&	01	21	56.861699	&	04	22	24.73436	&	2.7	\\
HD 8357	\dotfill	&	6454	&	0119+041	&	C	&	01	21	56.861699	&	04	22	24.73436	&	3.1	\\
RZ Cas	\dotfill	&	13133	&	0224+671	&	D	&	02	28	50.051459	&	67	21	03.02926	&	2.9	\\
B Per	\dotfill	&	20070	&	0355+508	&	O	&	03	59	29.747262	&	50	57	50.16151	&	3.0	\\
HD~283572	\dotfill	&	20388	&	0430+289	&	N	&	04	33	37.829860	&	29	05	55.47701	&	2.7	\\
T Tau N	\dotfill	&	20390	&	0409+229	&	N	&	04	12	43.666851	&	23	05	05.45299	&	4.2	\\
HD~37017	\dotfill	&	26233	&	0539$-$057	&	D	&	05	41	38.083384	&	$-$05	41	49.42839	&	2.0	\\
$\epsilon$ Ori	\dotfill	&	26311	&	0539$-$057	&	D	&	05	41	38.083384	&	$-$05	41	49.42839	&	4.7	\\
$\alpha$ Ori	\dotfill	&	27989	&	0529+075	&	C	&	05	32	38.998531	&	07	32	43.34586	&	5.6	\\
SV Cam	\dotfill	&	32015	&	0615+820	&	D	&	06	26	03.006188	&	82	02	25.56764	&	0.6	\\
HD 50896	\dotfill	&	33165	&	0646$-$306	&	C	&	06	48	14.096441	&	$-$30	44	19.65940	&	6.9	\\
R CMa	\dotfill	&	35487	&	0727$-$115	&	O	&	07	30	19.112472	&	$-$11	41	12.60048	&	5.4	\\
54 Cam	\dotfill	&	39348	&	0749+540	&	D	&	07	53	01.384573	&	53	52	59.63716	&	3.7	\\
TY Pyx	\dotfill	&	44164	&	0919$-$260	&	O	&	09	21	29.353874	&	$-$26	18	43.38604	&	5.1	\\
XY UMa	\dotfill	&	44998	&	0850+581	&	D	&	08	54	41.996385	&	57	57	29.93928	&	4.1	\\
IL Hya	\dotfill	&	46159	&	0919$-$260	&	O	&	09	21	29.353874	&	$-$26	18	43.38604	&	2.6	\\
DH~Leo	\dotfill	&	49018	&	0953+254	&	O	&	09	56	49.875361	&	25	15	16.04977	&	1.0	\\
HU Vir	\dotfill	&	59600	&	1145$-$071	&	C	&	11	47	51.554036	&	$-$07	24	41.14109	&	6.5	\\
DK Dra	\dotfill	&	59796	&	1053+704&	C	&       10    56     53.617492        &            70         11     45.91585 &       6.7   \\
RS CVn	\dotfill	&	64293	&	1315+346	&	C	&	13	17	36.494189	&	34	25	15.93266	&	2.1	\\
HR 5110	\dotfill	&	66257	&	1315+346	&	C	&	13	17	36.494189	&	34	25	15.93266	&	4.4	\\
RV Lib	\dotfill	&	71380	&	1430$-$178	&	C	&	14	32	57.690643	&	$-$18	01	35.24885	&	0.7	\\
$\delta$ Lib	\dotfill	&	73473	&	1511$-$100	&	C	&	15	13	44.893444	&	$-$10	12	00.26435	&	3.6	\\
AG Dra	\dotfill	&	78512	&	1642+690	&	D	&	16	42	07.848514	&	68	56	39.75640	&	4.4	\\
$\sigma^2$ CrB	\dotfill	&	79607	&	1611+343	&	C	&	16	13	41.064249	&	34	12	47.90909	&	0.4	\\
$\alpha$ Sco	\dotfill	&	80763	&	1622$-$253	&	O	&	16	25	46.891639	&	$-$25	27	38.32688	&	1.3	\\
WW Dra	\dotfill	&	81519	&	1637+574	&	D	&	16	38	13.456293	&	57	20	23.97918	&	3.4	\\
29 Dra	\dotfill	&	85852	&	1749+701	&	D	&	17	48	32.840231	&	70	05	50.76882	&	4.3	\\
Z Her	\dotfill	&	87965	&	1743+173	&	D	&	17	45	35.208181	&	17	20	01.42341	&	3.7	\\
9 Sgr	\dotfill	&	88469	&	1817$-$254	&	C	&	18	20	57.848685	&	$-$25	28	12.58456	&	4.0	\\
FR Sct	\dotfill	&	90115	&	1817$-$254	&	C	&	18	20	57.848685	&	$-$25	28	12.58456	&	12.8	\\
BY Dra	\dotfill	&	91009	&	1823+568	&	D	&	18	24	07.068372	&	56	51	01.49088	&	5.3	\\
HR 7275	\dotfill	&	94013	&	1954+513	&	D	&	19	55	42.738273	&	51	31	48.54623	&	7.3	\\
U Sge	\dotfill	&	94910	&	1923+210	&	C	&	19	25	59.605370	&	21	06	26.16218	&	2.3	\\
V444 Cyg	\dotfill	&	100214	&	2005+403	&	O	&	20	07	44.944851	&	40	29	48.60414	&	2.9	\\
HD 193793	\dotfill	&	100287	&	2005+403	&	O	&	20	07	44.944851	&	40	29	48.60414	&	4.1	\\
V729 Cyg	\dotfill	&	101341	&	2005+403	&	O	&	20	07	44.944851	&	40	29	48.60414	&	4.7	\\
HD 199178	\dotfill	&	103144	&	2037+511	&	D	&	20	38	37.034755	&	51	19	12.66269	&	7.4	\\
ER Vul	\dotfill	&	103833	&	2113+293	&	D	&	21	15	29.413455	&	29	33	38.36694	&	3.4	\\
VV Cep	\dotfill	&	108317	&	2229+695	&	D	&	22	30	36.469725	&	69	46	28.07698	&	7.0	\\
RT Lac	\dotfill	&	108728	&	2200+420	&	O	&	22	02	43.291377	&	42	16	39.97994	&	1.6	\\
AR Lac	\dotfill	&	109303	&	2200+420	&	O	&	22	02	43.291377	&	42	16	39.97994	&	3.6	\\
IM Peg	\dotfill	&	112997	&	2251+158	&	O	&	22	53	57.747932	&	16	08	53.56089	&	0.7	\\
SZ Psc	\dotfill	&	114639	&	2318+049	&	C	&	23	20	44.856598	&	05	13	49.95266	&	3.1	\\
$\lambda$ And	\dotfill	&	116584	&	2351+456	&	O	&	23	54	21.680266	&	45	53	04.23653	&	3.0	\\
HD 224085	\dotfill	&	117915	&	2337+264	&	O	&	23	40	29.029462	&	26	41	56.80485	&	3.8	\\
\enddata
\tablenotetext{a}{ICRF source category \citep{MA:98,IERS:99,FEY:04}: 
D = defining, C = candidate, O = other, N = new in ICRF-Ext. 1.}
\tablenotetext{b}{ICRF-Ext. 1 source positions \citep{IERS:99}.}\end{deluxetable}

\clearpage 

\begin{deluxetable}{lrrrrc}
\tabletypesize{\small}
\tablewidth{0pt}
\tablecaption{Radio Star Positions Estimated from the VLA+PT Data\label{POSITIONS}}
\tablehead{ \colhead{Star} & \colhead{\em Hipparcos} &  \colhead{$\alpha$ (J2000)} 
& \colhead{$\delta$ (J2000)} & \colhead{} & \colhead{} \\
\colhead{Name}  & \colhead{Number} & \colhead{(h m s)} & 
    \colhead{($^{\circ}$ $'$ $''$)} & \colhead{Epoch}  & \colhead{$N_{\rm obs}$ \tablenotemark{a}}}
\startdata
UV Psc	\dotfill	&	5980	&	01	16	55.1402	$\pm$	0.0012	($\pm$ 0.018$''$)	&	06	48	42.242	$\pm$	0.020	&	2003.6933	&	1/5	\\
HD 8357	\dotfill	&	6454	&	01	22	56.7799	$\pm$	0.0003	($\pm$ 0.004$''$)	&	07	25	10.126	$\pm$	0.006	&	2003.4356	&	6/6	\\
RZ Cas	\dotfill	&	13133	&	02	48	55.5126	$\pm$	0.0015	($\pm$ 0.008$''$)	&	69	38	03.563	$\pm$	0.010	&	2003.4356	&	5/6	\\
B Per	\dotfill	&	20070	&	04	18	14.6323	$\pm$	0.0005	($\pm$ 0.005$''$)	&	50	17	43.617	$\pm$	0.004	&	2003.4356	&	6/7	\\
HD~283572	\dotfill	&	20388	&	04	21	58.8519	$\pm$	0.0015	($\pm$ 0.019$''$)	&	28	18	06.379	$\pm$	0.019	&	2003.6933	&	3/5	\\
T Tau-N	\dotfill	&	20390	&	04	21	59.4364	$\pm$	0.0007	($\pm$ 0.009$''$)	&	19	32	06.394	$\pm$	0.011	&	2003.6933	&	5/5	\\
HD~37017	\dotfill	&	26233	&	05	35	21.8672	$\pm$	0.0006	($\pm$ 0.009$''$)	&	$-$04	29	39.013	$\pm$	0.009	&	2003.6933	&	6/7	\\
$\epsilon$ Ori	\dotfill	&	26311	&	05	36	12.8130	$\pm$	0.0005	($\pm$ 0.007$''$)	&	$-$01	12	06.924	$\pm$	0.012	&	2003.4356	&	6/7	\\
$\alpha$ Ori	\dotfill	&	27989	&	05	55	10.3097	$\pm$	0.0004	($\pm$ 0.005$''$)	&	07	24	25.461	$\pm$	0.012	&	2003.4356	&	7/7	\\
SV Cam	\dotfill	&	32015	&	06	41	19.1451	$\pm$	0.0061	($\pm$ 0.012$''$)	&	82	16	01.905	$\pm$	0.012	&	2003.4356	&	1/7	\\
HD 50896	\dotfill	&	33165	&	06	54	13.0405	$\pm$	0.0023	($\pm$ 0.032$''$)	&	$-$23	55	42.023	$\pm$	0.025	&	2003.6933	&	3/6	\\
R CMa	\dotfill	&	35487	&	07	19	28.2380	$\pm$	0.0003	($\pm$ 0.004$''$)	&	$-$16	23	43.564	$\pm$	0.007	&	2004.8000	&	3/5	\\
54 Cam	\dotfill	&	39348	&	08	02	35.7663	$\pm$	0.0004	($\pm$ 0.003$''$)	&	57	16	24.834	$\pm$	0.007	&	2003.6933	&	5/6	\\
TY Pyx	\dotfill	&	44164	&	08	59	42.7071	$\pm$	0.0008	($\pm$ 0.011$''$)	&	$-$27	48	58.911	$\pm$	0.012	&	2004.8000	&	3/5	\\
XY UMa	\dotfill	&	44998	&	09	09	55.9135	$\pm$	0.0029	($\pm$ 0.025$''$)	&	54	29	17.044	$\pm$	0.019	&	2003.6933	&	3/6	\\
IL Hya	\dotfill	&	46159	&	09	24	49.0001	$\pm$	0.0006	($\pm$ 0.008$''$)	&	$-$23	49	34.859	$\pm$	0.023	&	2004.8000	&	5/6	\\
DH~Leo	\dotfill	&	49018	&	10	00	01.6464	$\pm$	0.0008	($\pm$ 0.011$''$)	&	24	33	09.822	$\pm$	0.014	&	2003.6933	&	3/5	\\
HU Vir	\dotfill	&	59600	&	12	13	20.6889	$\pm$	0.0007	($\pm$ 0.010$''$)	&	$-$09	04	46.862	$\pm$	0.016	&	2004.8000	&	4/6	\\
DK Dra	\dotfill	&	59796	&	12	15	41.4825	$\pm$	0.0032	($\pm$ 0.014$''$)	&	72	33	04.221	$\pm$	0.009	&	2003.6933	&	5/6	\\
RS CVn	\dotfill	&	64293	&	13	10	36.8927	$\pm$	0.0005	($\pm$ 0.006$''$)	&	35	56	05.658	$\pm$	0.005	&	2003.4356	&	7/8	\\
HR 5110	\dotfill	&	66257	&	13	34	47.8330	$\pm$	0.0006	($\pm$ 0.007$''$)	&	37	10	56.655	$\pm$	0.003	&	2003.6933	&	4/5	\\
RV Lib	\dotfill	&	71380	&	14	35	48.4130	$\pm$	0.0005	($\pm$ 0.008$''$)	&	$-$18	02	11.598	$\pm$	0.011	&	2003.4356	&	5/5	\\
$\delta$ Lib	\dotfill	&	73473	&	15	00	58.3328	$\pm$	0.0005	($\pm$ 0.007$''$)	&	$-$08	31	08.248	$\pm$	0.011	&	2003.6933	&	3/6	\\
AG Dra	\dotfill	&	78512	&	16	01	41.0080	$\pm$	0.0030	($\pm$ 0.018$''$)	&	66	48	10.110	$\pm$	0.021	&	2003.4356	&	7/8	\\
$\sigma^2$ CrB	\dotfill	&	79607	&	16	14	40.7704	$\pm$	0.0007	($\pm$ 0.009$''$)	&	33	51	30.688	$\pm$	0.005	&	2003.6933	&	4/4	\\
$\alpha$ Sco	\dotfill	&	80763	&	16	29	24.4568	$\pm$	0.0005	($\pm$ 0.007$''$)	&	$-$26	25	55.286	$\pm$	0.014	&	2003.4356	&	5/5	\\
WW Dra	\dotfill	&	81519	&	16	39	03.9889	$\pm$	0.0014	($\pm$ 0.010$''$)	&	60	41	58.551	$\pm$	0.013	&	2003.6933	&	6/7	\\
29 Dra	\dotfill	&	85852	&	17	32	41.1473	$\pm$	0.0033	($\pm$ 0.013$''$)	&	74	13	38.567	$\pm$	0.015	&	2003.6933	&	7/7	\\
Z Her	\dotfill	&	87965	&	17	58	06.9733	$\pm$	0.0009	($\pm$ 0.013$''$)	&	15	08	22.179	$\pm$	0.013	&	2003.4356	&	6/6	\\
9 Sgr	\dotfill	&	88469	&	18	03	52.4445	$\pm$	0.0009	($\pm$ 0.012$''$)	&	$-$24	21	38.651	$\pm$	0.022	&	2003.6933	&	4/5	\\
FR Sct	\dotfill	&	90115	&	18	23	22.7919	$\pm$	0.0008	($\pm$ 0.011$''$)	&	$-$12	40	51.833	$\pm$	0.017	&	2003.4356	&	3/5	\\
BY Dra	\dotfill	&	91009	&	18	33	55.8399	$\pm$	0.0013	($\pm$ 0.012$''$)	&	51	43	07.724	$\pm$	0.014	&	2003.6933	&	7/7	\\
HR 7275	\dotfill	&	94013	&	19	08	25.7296	$\pm$	0.0016	($\pm$ 0.014$''$)	&	52	25	32.351	$\pm$	0.016	&	2004.8000	&	1/6	\\
U Sge	\dotfill	&	94910	&	19	18	48.4085	$\pm$	0.0002	($\pm$ 0.003$''$)	&	19	36	37.724	$\pm$	0.004	&	2003.4356	&	6/6	\\
V444 Cyg	\dotfill	&	100214	&	20	19	32.4209	$\pm$	0.0010	($\pm$ 0.011$''$)	&	38	43	53.954	$\pm$	0.008	&	2003.6933	&	4/6	\\
HD 193793 \tablenotemark{b}	\dotfill	&	100287	&	20	20	27.9752	$\pm$	0.0004	($\pm$ 0.004$''$)	&	43	51	16.271	$\pm$	0.002	&	2003.4356	&	5/6	\\
HD 193793	\dotfill	&	100287	&	20	20	27.9747	$\pm$	0.0004	($\pm$ 0.004$''$)	&	43	51	16.268	$\pm$	0.003	&	2004.8000	&	4/4	\\
V729 Cyg	\dotfill	&	101341	&	20	32	22.4221	$\pm$	0.0008	($\pm$ 0.009$''$)	&	41	18	18.919	$\pm$	0.011	&	2003.6933	&	5/5	\\
HD 199178	\dotfill	&	103144	&	20	53	53.6634	$\pm$	0.0014	($\pm$ 0.015$''$)	&	44	23	11.089	$\pm$	0.009	&	2003.4356	&	6/6	\\
ER Vul	\dotfill	&	103833	&	21	02	25.9309	$\pm$	0.0013	($\pm$ 0.018$''$)	&	27	48	26.485	$\pm$	0.017	&	2003.6933	&	6/6	\\
VV Cep	\dotfill	&	108317	&	21	56	39.1425	$\pm$	0.0019	($\pm$ 0.013$''$)	&	63	37	31.997	$\pm$	0.011	&	2003.4356	&	7/7	\\
RT Lac	\dotfill	&	108728	&	22	01	30.7601	$\pm$	0.0019	($\pm$ 0.021$''$)	&	43	53	25.734	$\pm$	0.012	&	2003.6933	&	7/7	\\
AR Lac	\dotfill	&	109303	&	22	08	40.8027	$\pm$	0.0003	($\pm$ 0.003$''$)	&	45	44	32.281	$\pm$	0.004	&	2003.4356	&	8/8	\\
IM Peg	\dotfill	&	112997	&	22	53	02.2589	$\pm$	0.0002	($\pm$ 0.003$''$)	&	16	50	28.168	$\pm$	0.003	&	2004.8000	&	6/6	\\
SZ Psc	\dotfill	&	114639	&	23	13	23.7901	$\pm$	0.0001	($\pm$ 0.001$''$)	&	02	40	31.689	$\pm$	0.004	&	2003.4356	&	6/6	\\
$\lambda$ And	\dotfill	&	116584	&	23	37	33.8999	$\pm$	0.0009	($\pm$ 0.010$''$)	&	46	27	27.808	$\pm$	0.007	&	2003.6933	&	6/6	\\
HD 224085	\dotfill	&	117915	&	23	55	04.2039	$\pm$	0.0004	($\pm$ 0.005$''$)	&	28	38	01.356	$\pm$	0.018	&	2003.4356	&	7/7	\\
\enddata
\tablenotetext{a}{Number of successful/total observations (scans).}
\tablenotetext{b}{HD 193793 was observed in experiments AF399a and AJ315.}
\end{deluxetable}

\clearpage 

\begin{deluxetable}{lrrrrrrrrr}
\tabletypesize{\small}
\tablewidth{0pt}
\tablecaption{Radio star position uncertainties and offsets from {\em Hipparcos}. \label{POS_ERRS_TAB}}
\tablehead{ 
\colhead{}     & \colhead{} & \multicolumn{2}{c}{Radio Errors} & 
\multicolumn{2}{c}{{\em Hipparcos} Errors \tablenotemark{a}} & 
\multicolumn{2}{c}{Combined Errors \tablenotemark{b}}  &  
\multicolumn{2}{c}{$\Delta_{ Hipp.-{\rm radio}}$} \\
\colhead{}     & \colhead{} & \multicolumn{2}{c}{(mas)} & 
\multicolumn{2}{c}{(mas)} & \multicolumn{2}{c}{(mas)}  & \multicolumn{2}{c}{(mas)} \\
\colhead{Star}     & \colhead{\em Hipparcos} & \multicolumn{2}{c}{\hrulefill} & 
\multicolumn{2}{c}{\hrulefill} & \multicolumn{2}{c}{\hrulefill}  & \multicolumn{2}{c}{\hrulefill} \\
\colhead{Name} & \colhead{Number} &  \colhead{$\alpha \cos \delta$ } 
& \colhead{$\delta$} & \colhead{$\alpha \cos \delta$ } & \colhead{$\delta$} 
& \colhead{$\alpha \cos \delta$ } & \colhead{$\delta$} 
& \colhead{$\alpha \cos \delta$ } & \colhead{$\delta$}
}
\startdata
UV Psc	\dotfill	&	5980	&	17.8	&	20.3	&	13.9	&	11.2	&	22.6	&	23.2	&	6.6	&	$-$51.4	\\
HD 8357	\dotfill	&	6454 &	4.4	&	5.8	&	12.0	&	8.1	&	12.7	&	10.0	&	2.2	&	7.4	\\
RZ Cas	\dotfill	&	13133	&	8.0	&	10.1	&	5.0	&	6.8	&	9.5	&	12.2	&	7.0	&	$-$1.3	\\
B Per	\dotfill	&	20070	&	5.1	&	4.4	&	14.3	&	11.5	&	15.2	&	12.3	&	25.1	&	$-$7.5	\\
HD~283572	\dotfill	&	20388	&	18.7	&	19.5	&	19.2	&	13.9	&	26.8	&	23.9	&	$-$22.8	&	23.2	\\
T Tau N	\dotfill	&	20390	&	10.0	&	11.0	&	23.0	&	19.8	&	25.0	&	22.6	&	35.2	&	$-$10.9	\\
HD~37017	\dotfill	&	26233	&	9.2	&	9.4	&	10.6	&	6.9	&	14.0	&	11.6	&	3.3	&	-3.6	\\
$\epsilon$ Ori	\dotfill	&	26311	&	6.9	&	12.1	&	9.8	&	5.4	&	12.0	&	13.3	&	12.1	&	10.0	\\
$\alpha$ Ori	\dotfill	&	27989	&	5.3	&	12.0	&	28.1	&	17.8	&	28.5	&	21.5	&	26.5	&	4.3	\\
SV Cam	\dotfill	&	32015	&	12.4	&	12.1	&	11.6	&	14.3	&	16.9	&	18.7	&	$-$5.1	&	$-$19.2	\\
HD 50896	\dotfill	&	33165	&	32.2	&	25.4	&	5.3	&	8.1	&	32.6	&	26.7	&	35.9	&	29.8	\\
R CMa	\dotfill	&	35487	&	3.6	&	6.6	&	9.4	&	9.6	&	10.0	&	11.7	&	8.3	&	26.3	\\
54 Cam	\dotfill	&	39348	&	2.8	&	6.7	&	9.5	&	7.7	&	9.9	&	10.2	&	$-$9.2	&	1.7	\\
TY Pyx	\dotfill	&	44164	&	10.6	&	12.3	&	6.4	&	7.5	&	12.3	&	14.4	&	2.8	&	3.6	\\
XY UMa	\dotfill	&	44998	&	25.9	&	19.1	&	21.1	&	14.4	&	33.5	&	23.9	&	34.6	&	$-$27.7	\\
IL Hya	\dotfill	&	46159	&	8.4	&	23.2	&	9.9	&	7.5	&	13.0	&	24.4	&	34.2	&	$-$11.2	\\
DH~Leo	\dotfill	&	49018	&	11.2	&	13.6	&	13.7	&	10.0	&	17.7	&	16.9	&	$-$49.5	&	$-$95.6	\\
HU Vir	\dotfill	&	59600	&	9.7	&	16.2	&	13.4	&	9.8	&	16.6	&	18.9	&	22.4	&	$-$20.9	\\
DK Dra	\dotfill	&	59796	&	8.5	&	9.1	&	6.8	&	6.2	&	10.9	&	11.0	&	0.0	&	$-$12.7	\\
RS CVn	\dotfill	&	64293	&	6.1	&	4.9	&	10.7	&	8.8	&	12.3	&	10.1	&	13.7	&	12.0	\\
HR 5110	\dotfill	&	66257	&	6.9	&	3.0	&	5.5	&	4.8	&	8.8	&	5.6	&	5.4	&	$-$6.9	\\
RV Lib	\dotfill	&	71380	&	7.6	&	11.3	&	21.5	&	17.3	&	22.8	&	20.7	&	4.8	&	$-$3.8	\\
$\delta$ Lib	\dotfill	&	73473	&	8.6	&	10.7	&	10.5	&	9.9	&	13.6	&	14.6	&	$-$18.5	&	41.5	\\
AG Dra	\dotfill	&	78512	&	17.9	&	21.0	&	11.2	&	13.2	&	21.2	&	24.8	&	11.3	&	10.3	\\
$\sigma^2$ CrB	\dotfill	&	79607	&	8.7	&	4.8	&	10.5	&	13.7	&	13.7	&	14.5	&	9.0	&	$-$4.6	\\
$\alpha$ Sco	\dotfill	&	80763	&	7.2	&	13.9	&	24.4	&	16.4	&	25.5	&	21.5	&	19.6	&	$-$3.2	\\
WW Dra	\dotfill	&	81519	&	10.0	&	12.6	&	18.9	&	18.9	&	21.4	&	22.8	&	34.2	&	18.6	\\
29 Dra	\dotfill	&	85852	&	13.0	&	15.2	&	10.6	&	11.5	&	16.8	&	19.1	&	$-$20.3	&	48.2	\\
Z Her	\dotfill	&	87965	&	13.0	&	13.0	&	8.4	&	7.8	&	15.5	&	15.1	&	19.5	&	$-$20.7	\\
9 Sgr	\dotfill	&	88469	&	12.5	&	21.6	&	14.7	&	8.8	&	19.3	&	23.3	&	7.2	&	16.7	\\
FR Sct	\dotfill	&	90115	&	11.0	&	17.0	&	19.8	&	14.1	&	22.6	&	22.0	&	$-$23.0	&	74.1	\\
BY Dra	\dotfill	&	91009	&	12.5	&	13.6	&	8.7	&	9.4	&	15.2	&	16.6	&	7.7	&	3.2	\\
HR 7275	\dotfill	&	94013	&	14.5	&	16.0	&	6.7	&	6.4	&	15.9	&	17.3	&	44.4	&	12.7	\\
U Sge	\dotfill	&	94910	&	2.9	&	3.6	&	6.6	&	7.1	&	7.2	&	8.0	&	$-$5.1	&	$-$4.6	\\
V444 Cyg	\dotfill	&	100214	&	13.1	&	8.1	&	8.4	&	8.4	&	15.6	&	11.7	&	$-$6.0	&	$-$4.0	\\
HD 193793	\dotfill	&	100287	&	3.9	&	2.6	&	7.1	&	6.0	&	8.1	&	6.5	&	$-$14.3	&	$-$4.6	\\
HD 193793	\dotfill	&	100287	&	4.1	&	2.3	&	7.9	&	6.7	&	8.9	&	7.0	&	$-$10.4	&	$-$4.9	\\
V729 Cyg	\dotfill	&	101341	&	8.7	&	10.9	&	33.6	&	29.5	&	34.7	&	31.4	&	26.4	&	29.8	\\
HD 199178	\dotfill	&	103144	&	14.8	&	8.8	&	9.4	&	7.5	&	17.5	&	11.5	&	$-$18.7	&	$-$8.1	\\
ER Vul	\dotfill	&	103833	&	17.6	&	17.4	&	7.0	&	7.0	&	19.0	&	18.8	&	8.1	&	$-$12.7	\\
VV Cep	\dotfill	&	108317	&	12.8	&	10.7	&	7.7	&	5.6	&	15.0	&	12.1	&	7.5	&	$-$3.9	\\
RT Lac	\dotfill	&	108728	&	20.8	&	12.0	&	10.3	&	11.0	&	23.2	&	16.3	&	1.1	&	$-$11.5	\\
AR Lac	\dotfill	&	109303	&	3.0	&	4.0	&	5.6	&	6.5	&	6.4	&	7.6	&	2.0	&	9.1	\\
IM Peg	\dotfill	&	112997	&	3.4	&	3.3	&	8.3	&	7.7	&	8.9	&	8.4	&	$-$5.6	&	$-$1.6	\\
SZ Psc	\dotfill	&	114639	&	1.4	&	4.3	&	14.9	&	9.9	&	15.0	&	10.8	&	5.4	&	$-$12.8	\\
$\lambda$ And	\dotfill	&	116584	&	9.5	&	6.5	&	4.0	&	6.2	&	10.3	&	9.0	&	$-$1.2	&	14.0	\\
HD 224085	\dotfill	&	117915	&	4.6	&	17.6	&	9.7	&	6.8	&	10.7	&	18.9	&	13.2	&	5.9	\\
\enddata
\tablenotetext{a}{{\em Hipparcos} uncertainties updated to the epoch of our observations using the 
{\em Hipparcos} proper motion uncertainties.}
\tablenotetext{b}{Combined uncertainties are the root-sum-square of our VLA+PT errors and 
the corresponding {\em Hipparcos} errors at epoch.}
\end{deluxetable}

\clearpage

\begin{deluxetable}{lrrrcr}
\tabletypesize{\small}
\tablewidth{0pt}
\tablecaption{Radio star proper motions. \label{PROP_MOT_TAB}}
\tablehead{
\colhead{Star} & \colhead{{\em Hipparcos}} &  \colhead{$\mu_{\alpha \cos\delta}$} & 
\colhead{$\mu_{\delta}$}  & \colhead{} & 
\colhead{$\Delta \tau$ \tablenotemark{b}}  \\
\colhead{Name}     & \colhead{Number}   & \colhead{(mas~yr$^{-1}$)} & 
\colhead{(mas~yr$^{-1}$)} &  \colhead{$N_{\rm pos}$ \tablenotemark{a}} & \colhead{(yr)} 
}
\startdata
UV Psc	\dotfill	&	5980	&	84.87	$\pm$	2.36	&	23.62	$\pm$	2.45	&	2	&	14.6893	\\
HD 8357	\dotfill	&	6454	&	96.17	$\pm$	1.17	&	234.07	$\pm$	1.20	&	3	&	18.3986	\\
RZ Cas	\dotfill	&	13133	&	3.20	$\pm$	1.51	&	36.55	$\pm$	1.60	&	5	&	17.1376	\\
B Per	\dotfill	&	20070	&	43.64	$\pm$	1.22	&	$-$58.02	$\pm$	1.21	&	7	&	21.0065	\\
HD~283572	\dotfill	&	20388	&	9.98	$\pm$	1.94	&	$-$29.00	$\pm$	1.90	&	2	&	18.6563	\\
T Tau N	\dotfill	&	20390	&	7.20	$\pm$	4.15	&	$-$4.84	$\pm$	5.17	&	2	&	2.7484	\\
HD~37017	\dotfill	&	26233	&	1.07	$\pm$	1.51	&	2.63	$\pm$	1.51	&	4	&	18.6563	\\
$\epsilon$ Ori	\dotfill	&	26311	&	$-$2.12	$\pm$	1.66	&	$-$0.91	$\pm$	1.78	&	3	&	17.2176	\\
$\alpha$ Ori	\dotfill	&	27989	&	23.98	$\pm$	1.04	&	10.07	$\pm$	1.15	&	4	&	21.0175	\\
SV Cam	\dotfill	&	32015	&	41.99	$\pm$	1.90	&	$-$150.02	$\pm$	1.81	&	2	&	17.1376	\\
HD 50896	\dotfill	&	33165	&	$-$3.70	$\pm$	2.06	&	6.58	$\pm$	1.94	&	3	&	20.0172	\\
R CMa	\dotfill	&	35487	&	166.19	$\pm$	1.64	&	$-$139.38	$\pm$	1.66	&	2	&	18.5820	\\
54 Cam	\dotfill	&	39348	&	$-$38.80	$\pm$	1.24	&	$-$56.95	$\pm$	1.39	&	4	&	21.2672	\\
TY Pyx	\dotfill	&	44164	&	$-$45.68	$\pm$	1.49	&	$-$46.24	$\pm$	1.52	&	3	&	21.1239	\\
XY UMa	\dotfill	&	44998	&	$-$50.19	$\pm$	2.24	&	$-$182.06	$\pm$	2.04	&	2	&	17.3953	\\
IL Hya	\dotfill	&	46159	&	$-$42.65	$\pm$	1.98	&	$-$30.58	$\pm$	2.39	&	2	&	15.7960	\\
DH~Leo	\dotfill	&	49018	&	$-$231.88	$\pm$	1.71	&	$-$31.78	$\pm$	1.77	&	2	&	18.6583	\\
HU Vir	\dotfill	&	59600	&	$-$13.22	$\pm$	1.99	&	1.33	$\pm$	2.15	&	2	&	15.7960	\\
DK Dra	\dotfill	&	58796	&	$-$9.55	$\pm$	1.79	&	$-$23.80	$\pm$	1.68	&	2	&	18.6583	\\
RS CVn	\dotfill	&	64293	&	$-$50.53	$\pm$	1.39	&	23.17	$\pm$	1.28	&	4	&	21.0145	\\
HR 5110	\dotfill	&	66257	&	85.40	$\pm$	1.07	&	$-$9.06	$\pm$	0.99	&	5	&	21.2752	\\
RV Lib	\dotfill	&	71380	&	$-$20.71	$\pm$	1.79	&	$-$18.24	$\pm$	1.86	&	2	&	17.2176	\\
$\delta$ Lib	\dotfill	&	73473	&	$-$66.16	$\pm$	1.51	&	$-$6.19	$\pm$	1.62	&	6	&	17.4753	\\
AG Dra	\dotfill	&	78512	&	$-$9.68	$\pm$	2.03	&	$-$7.61	$\pm$	2.13	&	2	&	17.2176	\\
$\sigma^2$ CrB	\dotfill	&	79607	&	$-$267.19	$\pm$	1.31	&	$-$86.86	$\pm$	1.22	&	4	&	21.2752	\\
$\alpha$ Sco	\dotfill	&	80763	&	$-$10.06	$\pm$	1.07	&	$-$23.81	$\pm$	1.24	&	3	&	21.0175	\\
WW Dra	\dotfill	&	81519	&	22.16	$\pm$	1.26	&	$-$60.40	$\pm$	1.32	&	3	&	20.0172	\\
29 Dra	\dotfill	&	85852	&	$-$66.85	$\pm$	1.89	&	33.88	$\pm$	1.92	&	2	&	17.4753	\\
Z Her	\dotfill	&	87965	&	$-$26.82	$\pm$	1.67	&	77.48	$\pm$	1.65	&	2	&	19.7596	\\
9 Sgr	\dotfill	&	88469	&	0.48	$\pm$	1.62	&	$-$3.94	$\pm$	1.85	&	2	&	20.0642	\\
FR Sct	\dotfill	&	90115	&	0.00	$\pm$	1.66	&	$-$2.88	$\pm$	1.74	&	2	&	19.8065	\\
BY Dra	\dotfill	&	91009	&	186.49	$\pm$	1.60	&	$-$325.12	$\pm$	1.65	&	2	&	20.0172	\\
HR 7275	\dotfill	&	94013	&	$-$102.73	$\pm$	1.81	&	$-$55.05	$\pm$	1.83	&	2	&	18.5820	\\
U Sge	\dotfill	&	94910	&	0.62	$\pm$	1.62	&	2.50	$\pm$	1.64	&	2	&	18.3986	\\
V444 Cyg	\dotfill	&	100214	&	$-$5.67	$\pm$	1.22	&	$-$7.62	$\pm$	1.13	&	3	&	20.0172	\\
HD 193793	\dotfill	&	100287	&	$-$4.72	$\pm$	0.66	&	$-$1.89	$\pm$	0.64	&	9	&	22.3819	\\
V729 Cyg	\dotfill	&	101341	&	$-$2.79	$\pm$	1.44	&	$-$5.35	$\pm$	1.48	&	4	&	20.0642	\\
HD 199178	\dotfill	&	103144	&	27.12	$\pm$	6.14	&	$-$1.03	$\pm$	3.51	&	3	&	2.4907	\\
ER Vul	\dotfill	&	103833	&	89.78	$\pm$	2.39	&	5.24	$\pm$	2.35	&	2	&	14.6893	\\
VV Cep	\dotfill	&	108317	&	$-$4.95	$\pm$	1.64	&	$-$2.27	$\pm$	1.61	&	2	&	19.8065	\\
RT Lac	\dotfill	&	108728	&	57.54	$\pm$	1.72	&	20.69	$\pm$	1.52	&	2	&	21.2672	\\
AR Lac	\dotfill	&	109303	&	$-$52.43	$\pm$	1.02	&	46.77	$\pm$	1.07	&	7	&	21.0095	\\
IM Peg	\dotfill	&	112997	&	$-$18.62	$\pm$	1.34	&	$-$27.76	$\pm$	4.22	&	2	&	3.8551	\\
SZ Psc	\dotfill	&	114639	&	18.64	$\pm$	1.41	&	27.60	$\pm$	1.43	&	3	&	19.8065	\\
$\lambda$ And	\dotfill	&	116584	&	158.77	$\pm$	1.47	&	$-$423.50	$\pm$	1.45	&	2	&	21.2752	\\
HD 224085	\dotfill	&	117915	&	576.20	$\pm$	1.45	&	34.27	$\pm$	1.67	&	2	&	21.0095	\\
\enddata
\tablenotetext{a}{Number of position measurements used in the weighted least-squares
fit to estimate the proper motion.}
\tablenotetext{b}{Time in years between first and last position epochs.}
\end{deluxetable}

\clearpage 

\begin{deluxetable}{lrrrrrrrrr}
\tabletypesize{\small}
\tablewidth{0pt}
\tablecaption{Radio star proper motion uncertainties and offsets from {\em Hipparcos}. \label{PM_ERRS_TAB}}
\tablehead{ 
\colhead{}     & \colhead{} & \multicolumn{2}{c}{Radio Errors} & 
\multicolumn{2}{c}{{\em Hipparcos} Errors} & 
\multicolumn{2}{c}{Combined Errors \tablenotemark{a}}  &
\multicolumn{2}{c}{$\Delta\mu_{ Hipp.-{\rm radio}}$} \\
\colhead{}     & \colhead{} & \multicolumn{2}{c}{(mas)} & 
\multicolumn{2}{c}{(mas)} & \multicolumn{2}{c}{(mas)}  & \multicolumn{2}{c}{(mas)} \\
\colhead{Star}     & \colhead{\em Hipparcos} & \multicolumn{2}{c}{\hrulefill} & 
\multicolumn{2}{c}{\hrulefill} & \multicolumn{2}{c}{\hrulefill}  & \multicolumn{2}{c}{\hrulefill} \\
\colhead{Name} & \colhead{Number} &  \colhead{$\mu_{\alpha \cos \delta}$ } 
& \colhead{$\mu_{\delta}$} & \colhead{$\mu_{\alpha \cos \delta}$} & \colhead{$\mu_{\delta}$} 
& \colhead{$\mu_{\alpha \cos \delta}$ } & \colhead{$\mu_{\delta}$} 
& \colhead{$\mu_{\alpha \cos \delta}$ } & \colhead{$\mu_{\delta}$}
}
\startdata
UV Psc	\dotfill	&	5980	&	2.36	&	2.45	&	1.14	&	0.92	&	2.63	&	2.62	&	0.03	&	4.39	\\
HD 8357	\dotfill	&	6454	&	1.17	&	1.20	&	0.98	&	0.66	&	1.53	&	1.37	&	1.78	&	3.34	\\
RZ Cas	\dotfill	&	13133	&	1.51	&	1.60	&	0.41	&	0.56	&	1.56	&	1.69	&	0.31	&	$-$0.92	\\
B Per	\dotfill	&	20070	&	1.22	&	1.21	&	1.17	&	0.94	&	1.69	&	1.53	&	$-$2.95	&	$-$1.59	\\
HD~283572	\dotfill	&	20388	&	1.94	&	1.90	&	1.57	&	1.14	&	2.50	&	2.22	&	2.46	&	$-$1.55	\\
T Tau N	\dotfill	&	20390	&	4.15	&	5.17	&	1.88	&	1.62	&	4.55	&	5.42	&	$-$8.25	&	7.64	\\
HD~37017	\dotfill	&	26233	&	1.51	&	1.51	&	0.86	&	0.56	&	1.73	&	1.61	&	$-$1.05	&	1.69	\\
$\epsilon$ Ori	\dotfill	&	26311	&	1.66	&	1.78	&	0.80	&	0.44	&	1.85	&	1.84	&	$-$3.61	&	0.15	\\
$\alpha$ Ori	\dotfill	&	27989	&	1.04	&	1.15	&	2.30	&	1.46	&	2.52	&	1.86	&	$-$3.35	&	$-$0.79	\\
SV Cam	\dotfill	&	32015	&	1.90	&	1.81	&	0.95	&	1.17	&	2.12	&	2.16	&	0.41	&	2.89	\\
HD 50896	\dotfill	&	33165	&	2.06	&	1.94	&	0.43	&	0.66	&	2.10	&	2.05	&	0.16	&	1.83	\\
R CMa	\dotfill	&	35487	&	1.64	&	1.66	&	0.69	&	0.71	&	1.78	&	1.80	&	0.82	&	$-$2.97	\\
54 Cam	\dotfill	&	39348	&	1.24	&	1.39	&	0.78	&	0.63	&	1.46	&	1.53	&	$-$0.52	&	2.13	\\
TY Pyx	\dotfill	&	44164	&	1.49	&	1.52	&	0.47	&	0.55	&	1.56	&	1.62	&	$-$1.69	&	$-$1.44	\\
XY UMa	\dotfill	&	44998	&	2.24	&	2.04	&	1.73	&	1.18	&	2.83	&	2.36	&	$-$2.11	&	2.61	\\
IL Hya	\dotfill	&	46159	&	1.98	&	2.39	&	0.73	&	0.55	&	2.11	&	2.46	&	$-$5.14	&	1.62	\\
DH~Leo	\dotfill	&	49018	&	1.71	&	1.77	&	1.11	&	0.81	&	2.04	&	1.95	&	2.49	&	4.33	\\
HU Vir	\dotfill	&	59600	&	1.99	&	2.15	&	0.99	&	0.72	&	2.22	&	2.27	&	$-$1.52	&	1.76	\\
DK Dra	\dotfill	&	59796	&	1.79	&	1.68	&	0.56	&	0.51	&	1.88	&	1.75	&	$-$0.35	&	1.31	\\
RS CVn	\dotfill	&	64293	&	1.39	&	1.28	&	0.88	&	0.72	&	1.64	&	1.47	&	$-$1.39	&	1.68	\\
HR 5110	\dotfill	&	66257	&	1.07	&	0.99	&	0.45	&	0.39	&	1.16	&	1.06	&	0.70	&	0.75	\\
RV Lib	\dotfill	&	71380	&	1.79	&	1.86	&	1.76	&	1.42	&	2.51	&	2.34	&	$-$0.17	&	0.62	\\
$\delta$ Lib	\dotfill	&	73473	&	1.51	&	1.62	&	0.86	&	0.81	&	1.74	&	1.81	&	0.04	&	$-$2.79	\\
AG Dra	\dotfill	&	78512	&	2.03	&	2.13	&	0.92	&	1.08	&	2.23	&	2.39	&	$-$3.50	&	$-$2.14	\\
$\sigma^2$ CrB	\dotfill	&	79607	&	1.31	&	1.22	&	0.86	&	1.12	&	1.57	&	1.66	&	$-$0.72	&	0.02	\\
$\alpha$ Sco	\dotfill	&	80763	&	1.07	&	1.24	&	2.00	&	1.34	&	2.27	&	1.83	&	0.10	&	$-$0.60	\\
WW Dra	\dotfill	&	81519	&	1.26	&	1.32	&	1.55	&	1.55	&	2.00	&	2.04	&	$-$2.60	&	$-$2.03	\\
29 Dra	\dotfill	&	85852	&	1.89	&	1.92	&	0.87	&	0.94	&	2.08	&	2.14	&	$-$0.03	&	$-$3.19	\\
Z Her	\dotfill	&	87965	&	1.67	&	1.65	&	0.69	&	0.64	&	1.81	&	1.77	&	$-$3.19	&	3.23	\\
9 Sgr	\dotfill	&	88469	&	1.62	&	1.85	&	1.20	&	0.72	&	2.02	&	1.99	&	$-$0.16	&	$-$2.23	\\
FR Sct	\dotfill	&	90115	&	1.66	&	1.74	&	1.62	&	1.15	&	2.32	&	2.09	&	1.76	&	$-$3.35	\\
BY Dra	\dotfill	&	91009	&	1.60	&	1.65	&	0.71	&	0.77	&	1.75	&	1.82	&	$-$0.13	&	$-$0.23	\\
HR 7275	\dotfill	&	94013	&	1.81	&	1.83	&	0.49	&	0.47	&	1.87	&	1.89	&	1.69	&	$-$0.78	\\
U Sge	\dotfill	&	94910	&	1.62	&	1.64	&	0.54	&	0.58	&	1.71	&	1.74	&	$-$2.32	&	$-$0.10	\\
V444 Cyg	\dotfill	&	100214	&	1.22	&	1.13	&	0.69	&	0.69	&	1.40	&	1.33	&	0.91	&	2.06	\\
HD 193793	\dotfill	&	100287	&	0.66	&	0.64	&	0.58	&	0.49	&	0.88	&	0.80	&	$-$2.00	&	$-$4.13	\\
V729 Cyg	\dotfill	&	101341	&	1.44	&	1.48	&	2.75	&	2.41	&	3.10	&	2.83	&	0.64	&	0.48	\\
HD 199178	\dotfill	&	103144	&	6.14	&	3.51	&	0.77	&	0.61	&	6.19	&	3.57	&	$-$2.11	&	$-$2.68	\\
ER Vul	\dotfill	&	103833	&	2.39	&	2.35	&	0.57	&	0.57	&	2.45	&	2.42	&	0.35	&	0.12	\\
VV Cep	\dotfill	&	108317	&	1.64	&	1.61	&	0.63	&	0.46	&	1.76	&	1.68	&	1.53	&	$-$0.86	\\
RT Lac	\dotfill	&	108728	&	1.72	&	1.52	&	0.84	&	0.90	&	1.91	&	1.77	&	$-$4.62	&	1.55	\\
AR Lac	\dotfill	&	109303	&	1.02	&	1.07	&	0.46	&	0.53	&	1.12	&	1.20	&	0.23	&	$-$0.46	\\
IM Peg	\dotfill	&	112997	&	1.34	&	4.22	&	0.61	&	0.57	&	1.47	&	4.26	&	0.05	&	$-$1.11	\\
SZ Psc	\dotfill	&	114639	&	1.41	&	1.43	&	1.22	&	0.81	&	1.87	&	1.64	&	2.35	&	$-$0.17	\\
$\lambda$ And	\dotfill	&	116584	&	1.47	&	1.45	&	0.33	&	0.51	&	1.51	&	1.54	&	0.53	&	1.54	\\
HD 224085	\dotfill	&	117915	&	1.45	&	1.67	&	0.79	&	0.56	&	1.65	&	1.76	&	$-$0.45	&	$-$2.04	\\
\enddata
\tablenotetext{a}{Combined uncertainties are the root-sum-square of our radio errors and 
the corresponding {\em Hipparcos} errors.}
\end{deluxetable}

\clearpage

\begin{deluxetable}{lrrrrrr}
\tabletypesize{\small}
\tablewidth{0pt}
\tablecaption{{\em Hipparcos}$-$radio data reference frame orientation and spin. \label{ROTATION}}
\tablehead{ \colhead{} & \colhead{$\epsilon_{x}$} & \colhead{$\epsilon_{y}$}  
                       & \colhead{$\epsilon_{z}$} & \colhead{$\omega_{x}$}
                       & \colhead{$\omega_{y}$}   & \colhead{$\omega_{z}$}  \\
            \colhead{} & \colhead{(mas)}     & \colhead{(mas)}
                       & \colhead{(mas)}     & \colhead{(mas~yr$^{-1}$)} 
                       & \colhead{(mas~yr$^{-1}$)}  & \colhead{(mas~yr$^{-1}$)}  } 
\startdata
all 46 stars  \dotfill            &  $-$0.4  &  0.1  & $-$3.2 &  0.55  &  0.02  & $-$0.41 \\
1$\sigma$ std.~error \dotfill &     2.6  &  2.6  &    2.9 &  0.34  &  0.36  &    0.37 \\
\tableline
33 stars \tablenotemark{a}     & $-$1.3 & 1.2  & $-$2.9  & 0.61 & $-$0.05 & $-$0.32 \\
1$\sigma$ std. error &    2.9 & 2.8  &  3.2    & 0.43 &   0.43  & 0.46 \\
\tableline
31 stars \tablenotemark{b}     & $-$1.7 & 0.9  & $-$2.4  & 0.62 & $-$0.01 & $-$0.30 \\
1$\sigma$ std. error &    3.1 & 2.9  &  3.3    & 0.43 &   0.43  & 0.46 \\
\enddata
\tablenotetext{a}{Solution excluding 13 stars with {\em Hipparcos} multiplicity flags.}
\tablenotetext{b}{Solution excluding 13 {\em Hipparcos} multiples plus T Tau N and RZ Cas.}
\end{deluxetable}


\begin{thebibliography}{}

\bibitem[Boboltz et al.(2003)]{BFJCdZG:03}
Boboltz, D.~A., Fey, A.~L., Johnston, K.~J., Claussen, M.~J., 
de Vegt, C., Zacharias, N., \& Gaume, R.~A.\ 2003, \aj, 126, 484 

\bibitem[Claussen et al.(1999)]{CBSU:99}
Claussen, M.~J., Beresford, R., Sowinski, K. \& Ulvestad, J.~S.\ 1999, 
\baas, 195, 83.05 

\bibitem[Fey et al.(2004)]{FEY:04}
Fey, A. L., et al. 2004, \aj, 127, 3587

\bibitem[Fey et al.(2006)]{FBGJGT:06} 
Fey, A.~L., Boboltz, D.~A., Gaume, R.~A., Johnston, K.~J., 
Garrington, S.~T., \& Thomasson, P.\ 2006, \aj, 131, 1084 
 
\bibitem[Gambis(1999)]{IERS:99}
Gambis, D. ed. 1999, 1998 IERS Annual Report, Observatoire de Paris, VI, 87 

\bibitem[Hartkopf et al.(2001b)]{HMWM:01}
Hartkopf, W.~I., Mason, B.~D., Wycoff, G.~L., \& McAlister, H.~A.\ 2001b, 
Fourth Catalog of Interferometric Measurements of Binary Stars, 
http://www.ad.usno.navy.mil/wds/int4.html

\bibitem[Johnston et al.(1985)]{JDFW:85}
Johnston, K. J., de Vegt, C., Florkowski, D. R., \& Wade, C. M.1985,
\aj, 90, 2390

\bibitem[Johnston et al.(2003)]{JDG:03} 
Johnston, K. J., de Vegt, C., \& Gaume, R. A. 2003, \aj, 125, 3252

\bibitem[Johnston et al.(2004)]{JFGCH:04} 
Johnston, K.~J., Fey, A.~L., Gaume, R.~A., Claussen, M.~J., \& 
Hummel, C.~A.\ 2004, \aj, 128, 822 

\bibitem[Kovalevsky et al.(1997)]{KOVAL:97}
Kovalevsky, J., et al. 1997, \aap, 323, 620

\bibitem[Lestrade et al.(1999)]{LPJPRTRG:99}
Lestrade, J.-F, Preston, R.~A., Jones, D.~L., Phillips, R.~B., Rogers, A.~E.~E., 
Titus, M.~A., Rioja, M.~J., \& Gabuzda, D.~C.\ 1999, \aap, 344, 1014

\bibitem[Ma et al.(1998)]{MA:98}
Ma, C., et al. 1998, \aj, 116, 516

\bibitem[Mason et al.(2001)]{MWHDW:01} 
Mason, B.~D., Wycoff, G.~L., Hartkopf, W.~I., Douglass, G.~G., \& 
Worley, C.~E.\ 2001, \aj, 122, 3466 

\bibitem[Perryman et al.(1997)]{PERRYMAN:97}
Perryman M.~A.~C., et al.\ 1997, \aap, 323, L49 

\bibitem[Walter \& Sovers(2000)]{WS:00} 
Walter, H.~G., \& Sovers, O.~J.\ 2000, 
Astrometry of fundamental catalogues : the evolution from optical to 
radio reference frames, edited by Hans G.~Walter and Ojars 
J.~Sovers.~ Berlin, Heidelberg: Springer-Verlag 


\end{thebibliography}
\end{document}